\documentclass{jfp1}

\usepackage{combelow}
\usepackage{listings}

\usepackage{pgf}
\usepackage{tikz}
\usetikzlibrary{arrows,automata}

\newcommand\bcmdtab{\noindent\bgroup\tabcolsep=0pt%
  \begin{tabular}{@{}p{10pc}@{}p{20pc}@{}}}
\newcommand\ecmdtab{\end{tabular}\egroup}

\title[KompicsTesting]
      {KompicsTesting - Unit Testing Event Streams}

 \author[I. W. Ubah et al.]
        {IFEANYI W. UBAH, LARS KROLL, ALEXANDRU A. ORMENI\cb{S}AN, SEIF HARIDI\\
         Department of Software and Computer Systems, School of ICT\\
         KTH Royal Institute of Technology, Stockholm, Sweden
         }

\jdate{May 2017}
\pubyear{2017}
\pagerange{\pageref{firstpage}--\pageref{lastpage}}

\begin{document}

\label{firstpage}

\maketitle

\begin{abstract}
In this paper we present KompicsTesting, a framework for unit testing components in the Kompics component model.
Components in Kompics are event-driven entities which communicate asynchronously solely by message passing.
Similar to actors in the actor model, they do not share their internal state in message-passing, making them less prone to errors, compared to other models of concurrency using shared state.
However, they are neither immune to simpler logical and specification errors nor errors such as dataraces that stem from nondeterminism.
As a result, there exists a need for tools that enable rapid and iterative development and testing of message passing components in general, in a manner similar to the xUnit frameworks for functions and modular segments code. These frameworks work in an imperative manner, ill suited for testing message-passing components given that the behavior of such components are encoded in the streams of messages that they send and receive.
In this work, we present a theoretical framework for describing and verifying the behavior of message-passing components, independent of the model and framework implementation, in a manner similar to describing a stream of characters using regular expressions. We show how this approach can be used to perform both black box and white box testing of components and illustrate its feasibility through the design and implementation a prototype based on this approach, KompicsTesting.
\end{abstract}

\tableofcontents

\lstset{
  mathescape,
  captionpos=b,                    
  extendedchars=true,              
  frame=single,                    
  language=Java,                 
  keywordstyle=\bf,
  showspaces=false,
  showstringspaces=false,          
  showtabs=false,                  
  tabsize=2,                       
  stepnumber=1,
  numbers=left,
  numbersep=5pt,                   
  numberstyle=\tiny\color{gray},
  morekeywords={repeat, body, end, allow, disallow, drop, expect, inspect, fault, blockExpect, trigger, unordered, either, or},
  literate={->}{$\rightarrow$}{2}
           {epsilon}{$\varepsilon$}{1}
           {num}{\textit{n}}{1}
           {ev_}{\textit{e}}{1}
           {ac_}{\textit{a}}{1}
           {faultType}{\textit{f}}{1}
           {resolveAction}{\textit{r}}{1}
           {cb}{\(\alpha\)}{1}
           {a1}{$a_1$}{2}
           {a2}{$a_2$}{2}
           {a3}{$a_3$}{2}
           {a4}{$a_4$}{2}
           {a5}{$a_5$}{2}
           {a_n}{$a_n$}{2}
           {a_i}{$a_i$}{2}
           {e0}{$e_0$}{2}
           {e1}{$e_1$}{2}
           {e2}{$e_2$}{2}
           {e3}{$e_3$}{2}
           {e4}{$e_4$}{2}
           {e5}{$e_5$}{2}
           {e_n}{$e_n$}{2}
           {e_i}{$e_i$}{2}
}

\section{Introduction}
Companies spend a large amount of resources in a quest to detect and fix errors in their software systems. Such errors are usually introduced into the systems either at development time, due to mistakes made by programmers, or during system requirements specification, and may untimely present themselves, resulting in failures with possibly severe consequences.
The chances of introducing errors rapidly increase with the complexity of the system and as more complex systems are developed to take prominent roles in our lives, it becomes increasingly important that as many errors as possible are found and fixed as early as possible in the development process.

Message-passing systems are inherently complex and this complexity is amplified when processes work independent of each other in a distributed setting, due to the inherent asynchrony and possibility of partial failure where subsets of processes may fail at any time during execution.

One way of finding and detecting errors in a system is by testing, that is the execution of a system with the sole purpose of finding errors. This can be performed on various levels depending on how much of, and at what granularity the system is to be exercised.
At the lowest level is unit testing \cite{IEEESEVocab2010}, focusing on the verification of the smallest testable entities in the system, usually a single function, class or module in object oriented programming languages, or an actor, process, or component in a message-passing system.

However, when unit testing an entity in a message passing system (which we refer to such an entity simply as a component from hereon) however, the task of specifying meaningful tests become convoluted if at all feasible.
Most tools available for unit testing only allow the tester to describe tests in a low-level manner, for example, using mechanisms for sequentially making assertions on the return values of functions.
While this way of testing may prove effective for the intended languages or frameworks, they are not sufficient for verifying the behavior of a component as this can not be done properly without examining the sequences of messages sent and received by the component.

Our approach enables the tester to describe the expected behavior of the component as a stream or sequence of messages and their directions.
For example, a tester may deem the behavior of a component under test (CUT) to be correct if the component responds with a message \(m_2\) on receiving a message \(m_1\) and subsequently responds with one or more messages \(m_4\) on receiving \(m_3\).
This could be specified as \\ \((in(m_1), out(m_2), in(m_3), out(m_4), out(m_4)^*)\) where \emph{in} and \emph{out} denote incoming and outgoing messages from the perspective of the CUT and $^*$ is the Kleene closure operator denoting zero or more occurrences of the message \(out(m_4)\).
Such a specification of sequences form a language. If we assign each message and direction pair the values \(e_1, e_2, e_3, e_4\) respectively and call each such pair an event, then the language of the specification is thus described by the regular expression \(e_1e_2e_3e_4^*\) and the alphabet of the described language are the event symbols \(\{e_1, e_2, e_3, e_4\}\).

Such a specification can subsequently be converted to an automaton by a framework implementation and used to verify the CUT by running an instance and verifying that the events observed conform to the specification.
Thus, if the automaton ends up in an accepting state when all events have been observed, the test case passes.

This paper presents a theoretical framework for unit testing message-passing systems using this approach as well as KompicsTesting, an implementation of the framework for the Kompics component model.
Section \ref{sec:kompics} presents the Kompics component model while sections \ref{sec:formalgrammar} to \ref{sec:cfg} give a brief background on formal languages and finite automata needed to follow the paper.
We then describe the aforementioned theoretical framework in section \ref{sec:approach}, followed by a description of the design and implementation of KompicsTesting in section \ref{sec:kompicstesting}. This is followed by a discussion of related work (section \ref{sec:relatedwork}) and concluding remarks (section \ref{sec:conclusion}).

\subsection{The Kompics Component Model}
\label{sec:kompics}
Kompics is a component based, message-passing model for building distributed systems.
Components in Kompics are event-driven entities that communicate by exchanging messages, in the form of events, with each other.
Events are simply data-carrying objects in the system.
Components provide communication interfaces via bidirectional ports and are connected to each other via channels binding any two ports.
The following sections describe the key primitives and concepts used in the Kompics model; more details can be found in \cite{Arad2013a}.

\subsubsection{Ports}
Ports in Kompics embody the interface between a component and its environment.
They are \emph{bidirectional} entities through which events are sent to and received from a component.
Through ports, Kompics provides a type system for the events exchanged in the system. Unlike systems like Akka \cite{Wyatt:2013:AC:2663429} and Erlang \cite{Armstrong2003} where any message may be addressed to an actor or component, ports define which events may go in and out of a component.

Each port has two directions which we label \emph{positive} and \emph{negative}, as well as a \emph{port type} which declares a specific set of event types that are allowed to pass through it in each direction.
We denote a port \(\alpha = (p_\alpha, n_\alpha)\) where \(p_\alpha\) is the positive direction and \(n_\alpha\) is its negative direction, and say that a port \(\alpha\) allows an event \(e\) in some direction \(d_\alpha\) if its port type declares the type of \(e\) in direction \(d_\alpha\).

As communication interfaces, we can think of a port \(\alpha = (p_\alpha, n_\alpha)\) as a service interface and associate requests with its negative side \(n_\alpha\) and responses with its positive side \(p_\alpha\).
Thus a component that produces service \(\alpha\), declares \(\alpha\) so that requests events are received, incoming from \(n_\alpha\) while responses events are outgoing from \(p_\alpha\).
We also say that the component declaring \(\alpha\) in this manner \emph{provides} the port.
Conversely, a component that consumes service \(\alpha\), declares 
\(\alpha\) so that responses are incoming from \(p_\alpha\) while requests are outgoing from \(n_\alpha\). We say that the component \emph{requires} the port \(\alpha\).

We again, label the two sides of a port from the perspective of a particular component declaring the port by saying that the side of the port emitting incoming events to the component is the \emph{inside} port while the side emitting outgoing events is the \emph{outside} port.
Note that for a component that provides the port \(\alpha\), the inside and outside ports are \(n_\alpha\) and \(p_\alpha\) respectively while for the component that requires \(\alpha\), they are conversely \(p_\alpha\) and \(n_\alpha\) respectively

Finally, we say a component \emph{triggers} an event on a port, if it sends an event through that port --- this is typically done from the inside, going out, except when interfacting component code with traditional thread-based libraries.

\subsubsection{Channels}
Channels create connections between two components via their declared ports.
They can be thought of as bidirectional communication pipes that carry events from one port to another.
Connections are only possible for any two ports of the same port type that have opposite directions, that is a channel can be created between two ports \(\alpha\) and \(\beta\) by connecting \(p_\alpha\) to \(n_\beta\) or \(n_\alpha\) to \(p_\beta\).
Events are fowarded through channels in first in, first out (FIFO) order, that is they are delivered to the destination port in the order that they were triggered at the source port.
An event triggered on a port is broadcast on all channels connected to that port, thus Kompics does not provide a mechanism for addressing events to specific components in the system.
On arrival of events  at their destination ports, they are queued up on that port until the component that declared the port is scheduled to execute that particular event on that port.

\subsubsection{Event Handlers}
An event handler, or \emph{handler}, in Kompics is a user-defined function of a component.
A handler accepts events of a particular type and any of its subtypes (in the strongly typed programming language sense).
Handlers are registered on ports and as functions are executed whenever the component receives an event on an event that is accepted by any registered handlers.
We say that a registered handler for a given port is \emph{subscribed} to the port.

Kompics guarantees the sequential execution of the handlers of a single component, preventing the need for internal state synchronization between handlers. Handlers belonging to different components on the other hand, may be scheduled and executed concurrently.

\subsubsection{Components}
Components in Kompics are reactive entities which communicate asynchronously with each other by exchanging messages.
Similar to actor based systems, a component has some internal state associated with it as well as an event queue at its declared ports.
It also has a set of event handlers as described, \emph{subscribed} on its declared ports, which are executed whenever some accepted event is received on that port.
Handlers are the means by which a component updates its internal state.

A component can be encapsulated within another component, using parent-child relationships that form a component heirarchy, with a single \emph{Main} component that is initially started at runtime.
The relationship between a component and its child components enables a flexible architecture for managing system complexity as well as the delegation of configuration of a component to the parent.
For example, on creating and starting a component, the sub-components or child components are recursively created and started.
It is the job of such a parent component to boostrap its child components by say, setting up their communication channels.

\subsection{Formal Languages and Grammars}
\label{sec:formalgrammar}

A formal language \cite{Mateescu:1997:FLI:267846.267847} is a set of sequences of symbols formed together according to some specified rule.
A language has an alphabet, which is the set of allowed symbols that may be used to form valid sequences of that language.
As an example, the set \(\{ab,ac\}\) is a language containing exactly two sequences, the alphabet is the set of \emph{characters} \(\{a, b, c\}\).

It is rarely feasible to describe a language as we just did by listing out all sequences that are members of that language, nor is it particularly desireable to verbally describe all possible languages.
Instead, one may simply give a \textit{formation rule} that describes the intended language.
This rule may then be used to either generate sequences that belongs to the described language or check that a given sequence belongs to the language.
Such a rule may be a grammar \cite{chomsky1956three} or an automaton \cite{hopcroftullman79} and several of these exist for a variety of classes of languages.
We discuss finite automata as well as regular and context-free grammars.

We start by defining the Kleene closure, or closure, on a set of symbols. Let \(\Sigma\) be a finite set of symbols. The closure \(\Sigma^*\) on \(\Sigma\) is the set of sequences that can formed taking any number of symbols in \(\Sigma\), with repetitions allowed.
For example if \(\Sigma = \{a,b\}\), the closure \(\Sigma^* = \{a, b, aa, ab, aba,...\}\).
The closure on any set always includes the empty symbol \(\varepsilon\) since one may choose to take zero number of symbols.

A grammar \(G\) is formally described as a four-tuple \((N, \Sigma, P, S)\) where \(N\) is a set of \emph{nonterminal} symbols, \(\Sigma\) is the alphabet containing \emph{terminal} symbols, \(P\) a set of \emph{productions} and \(S \in N\) a start symbol \cite{hopcroftullman79}.
A production \(p \in P\) is of the form \((\Sigma \enskip \cup \enskip N)^*N(\Sigma \enskip \cup \enskip N)^* \to (\Sigma \enskip \cup \enskip N)^*\) where \(^*\) is the closure operator --- hence the right hand side may be empty.
Thus a production may have at least one nonterminal symbol and any number of terminal and nonterminal symbols on the left-hand side of the arrow while the right-hand side may contain any number of terminals and nonterminals.

If the grammar \(G\) generates exactly the set of sequences that make up some language \(\mathrm{L}\), then we say that the grammar describes\(\mathrm{L}\) and write \(L(G) = \mathrm{L}\).
To generate a sequence in \(\mathrm{L}\) using \(G\), we start with the start symbol \(S\) and expand using the right-hand side of any production of \(S\), replacing the expanded symbol with its right-hand side.
We repeatedly do this until no nonterminal symbols are left in the expanded sequence.

As an example, consider the simple grammar \(G\) whose productions are shown in listing \ref{lst:simplegrammar}. Nonterminal symbols are denoted with capital letters while terminal symbols are denoted by small letters. So \(N = \{A,B\}\) and alphabet \(\Sigma = \{a,b,c,d\}\), \(S\) is the start symbol.
We begin with the start symbol and expand the sequence using the first production. Thus we replace \(S\) with the sequence \(ABc\).
Next, we replace the next nonterminal \(A\) with its right-hand side \(a\) leaving the new sequence \(aBc\).
Finally, we do the same for \(B\) to retrieve the final sequence containing no terminals \(abc\).
Note that we could have started by expanding \(S\) using the second production which would have resulted in the string \(bcd\).

\begin{lstlisting}[label=lst:simplegrammar, float=t!, caption={A Simple Grammar.}]
S -> ABc
S -> Bcd
A -> a
B -> b
\end{lstlisting}

\subsection{Regular Grammars and Expressions}
\label{sec:regulargrammar}
Regular grammars restrict the types of productions they contain. Here, the left-hand side must contain exactly one nonterminal symbol while the right-hand side may contain a single terminal symbol followed by a nonterminal symbol, or a nonterminal symbol followed by a terminal symbol. The right-hand side may also be empty.
A language described by a regular grammar is called a \emph{regular} language.
We can forego the use of regular grammars inplace of an equivalent \emph{regular expressions} \cite{kleene1951representation} notation for describing regular languages in a more concise manner.

A single regular expression describes a language that is regular.
The basis for regular expressions are the symbols of the alphabet \(\Sigma\), such that if \(a \in \Sigma\) then \(L(a) = \{a\}\), that is the language described by a terminal symbol of an alphabet consists of the single symbol.
Regular expressions can be built from smaller regular expressions using the following operations on that are closed on regular languages \cite{hopcroftullman79}.

\begin{description}
  \item[Union] The union of two languages \(L\) and \(M\) denoted \(L \cup M\) is the language containing all sequences that are either in \(L\) or in \(M\) or both. For example, if \(L = \{ab\}\) and \(M = \{ab, cd\}\), then the union \(L \cup M\) is the language \(\{ab, cd\}\). Hence if \(R\) and \(S\) are regular expressions, then \(R|S\) is also a regular expresion describing the language \(L(R) \cup L(S)\).
  \item[Concatenation] The concatenation of two language \(L\) and \(M\) denoted \(LM\) is the language containing all sequences that are formed by taking a sequence in \(L\) and appending a sequence in \(M\) to it. Using the previous example, the concatenation of both languages \(LM = \{abcd\}\). Hence if \(R\) and \(S\) are regular expressions, then \(RS\) is also a regular expression describing the language \(L(R)L(S)\).
  \item[Kleene Closure] The closure on a language \(L\), denoted \(L^*\) is the language containing all sequences formed by taking any number of sequences, with repetitions allowed, from \(L\) and concatenating them.
  Using the previous example, the closure \(L^*\) on \(L\) is the language \(\{ab, abab, ababab, ...\}\). Hence if \(R\) is a regular expression \(R^*\) is also a regular expression.
\end{description}

\subsection{Finite Automata}
A finite state automaton (from hereon automaton) is an abstract machine made up of states and transitions which are labelled with an input symbol and connect any two states \cite{hopcroftullman79}. An automaton can be in any such state at a given time, changing states by following transitions and consuming the input associated with that transition.
An automaton can either be a \emph{deterministic} finite automaton (DFA) or a \emph{nondeterministic} finite automaton (NFA).
A DFA has exactly one destination state from any given state and input symbol while an NFA may have several possible destination states. As a result, an NFA may be in several states at once.

\subsubsection{Deterministic Finite Automata}
\label{sec:dfa}
Formally, a DFA \(M\) is defined as the five-tuple \((Q, \enskip \Sigma, \enskip \delta, q_0, F)\) where \(Q\) is a finite set of states, \(\Sigma\) is the set of input symbols, \(\delta\) is a transition function that takes an input symbol \(i \in \Sigma\) and a state \(q_i \in Q\) and maps them to a new state \(q_n \in Q\), \(q_0\) is a start state which is the initial state of the automaton, \(F \subseteq Q\) is a set of accepting states \cite{hopcroftullman79}.

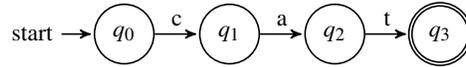
\begin{figure}
\begin{tikzpicture}[->,>=stealth',shorten >=1pt,auto,node distance=1.4cm,
                    semithick,label=foo]

  \node[initial,state]           (0)              {$q_0$};
  \node[state]                   (1) [right of=0] {$q_1$};
  \node[state]                   (2) [right of=1] {$q_2$};
  \node[state,accepting]         (3) [right of=2] {$q_3$};

  \path   (0) edge              node  {c} (1)
          (1) edge              node  {a} (2)
          (2) edge              node  {t} (3);
\end{tikzpicture}
\caption{A DFA recognizing the sequence ``cat''.}
\label{fig:dfaexample}
\end{figure}

Instead of explicitly listing out the mappings of the transition function \(\delta\) of a finite automaton, we can visually describe it using a state diagram as shown in figure \ref{fig:dfaexample} .
Here, \(Q = \{q_0, q_1, q_2, q_3\}\), \(\Sigma = \{a,c,t\}\), \(F = \{q_3\}\). Transitions are arrows labelled by an input symbol. The start state \(q_0\) has no incoming transitions while final states, here \(q_3\), are depicted by a double-bordered circle.

A DFA can be used to recognize a sequence of symbols. Given a sequence, it can be simulated such that at every given state, it consumes the next symbol and follows a transition labeled with that symbol to the next state.
When the sequence is completely consumed, if the DFA is in an accepting state, then we say that it \emph{recognizes} or \emph{accepts} the sequence otherwise we say that it \emph{rejects} the sequence.
The language \(L(M)\) of a DFA \(M\) is the set of all sequences recognized by \(M\), we also say that \(M\) describes \(L(M)\).

As an example, listing \ref{fig:dfaexample} shows a DFA that recognizes the sequence ``cat''. It starts at state \(q_0\) and consumes the first symbol `c', from \(q_0\) it follows the transition labeled with this symbol to \(q_1\), then it consumes the next symbol `a' and so on until no more symbols are left.
Since it ends up in state \(q_3\), we say that it recognizes the input sequence ``cat''.

Note that we have not specified transitions for every other symbol at each state, and the DFA must have exactly one valid transition for each state and input pair. We only show those transitions that lead to an accepting state and imply that all other non-specified transitions lead to an implicit error state such that the automaton rejects the sequence.

Finally we note that, like for regular grammars and regular expressions, the language \(L(M)\) of a DFA \(M\) is also a regular language \cite{hopcroftullman79}. Thus if a language can be described by a regular grammar, then it can also be described by a DFA. In fact, every DFA can be converted into an equivalent regular expression and vice-versa.

\subsubsection{Nondeterministic Finite Automata}
As mentioned previously, a NFA can be in several states at the same time since it may have multiple transitions for a given input symbol and state pair. The automaton is said to guess its next state by following all possible transitions.

Formally, an NFA \(M\) is defined as the five-tuple \((Q, \enskip \Sigma, \enskip \delta, \enskip q_0, \enskip F)\), where \(Q\) is a finite set of states, \(\Sigma\) is a finite set of input symbols, \(q_0 \in Q\) is the start state, \(F \subseteq Q\) is a set of accepting states and \(\delta\) is a transition function that takes a state \(q_i \in Q\) and an input symbol \(i \in \Sigma\) and returns a \emph{subset} of \(Q\) \cite{hopcroftullman79}. Thus, the transition function \(\delta\) alone marks the only difference between an NFA and a DFA.
NFAs describe the same class of languages as DFAs, in fact, any NFA can be converted to an equivalent DFA using the subset construction \cite{Rabin:1959:FAD:1661907.1661909}. Conversely, note that every DFA is an NFA such that the transition function always returns a singleton set.

Finally, an NFA with epsilon transitions, or \(\varepsilon\)-NFA is an extended NFA with the single additional ability to allow a transition without consuming any input string.
Thus an \(\varepsilon\)-NFA can have unlabeled transitions, called \(\varepsilon\)-transitions, allowing it to make spontaneous a transition to the next state. In a state diagram, such labels are marked with the special \(\varepsilon\) symbol only as a visual convenience, the symbol does not belong to the input alphabet of the automaton.
\(\varepsilon\)-NFAs do not extend the class of languages defined by NFAs or DFAs or regular grammars, they all define regular languages. In fact, any given \(\varepsilon\)-NFA can be converted to an equivalent \(NFA\).
However, they offer more illustrative and programming convenience and will be used later in this paper.

\subsubsection{Concatenating Finite Automata}
Since the described finite automata all define the same class of language as regular grammar and expressions do, the union, concatenation and closure operations can also be performed on any such automaton.
Here we describe the construction of an automaton \(MN\) from the concatenation of any two automata \(M\) and \(N\) as presented in \cite{hopcroftullman79}.
This operation is implied throughout this paper and as such is presented here for reference.
A more specific description for the union and closure operations on automata is illustrated in sections \ref{sec:union} and \ref{sec:closure}

To construct an automaton \(MN = (Q, \enskip \Sigma, \enskip \delta, \enskip q_0, \enskip  F)\) from two automata \(M\) and \(N\) where \(M = (Q_M, \enskip \Sigma_M, \enskip \delta_M, \enskip q_{0M}, \enskip  F_M)\) and \(N = (Q_N, \enskip \Sigma_N, \enskip \delta_N, \enskip q_{0N}, \enskip  F_N)\), We set \(Q = Q_M \cup Q_N\), next we set the start state of the first automaton \(q_{0M}\) as the start state \(q_0\) of \(MN\) and the accepting states \(F_N\) of the second automaton as the accepting states \(F\) of \(MN\).
The transition function \(\delta\) uses the mappings from \(\delta_M\) and \(\delta_N\) with additional \(\varepsilon\)-transitions from each final state \(q \in F_M\) in the first automaton to the start state \(q_{0N}\) of the second automaton.

The idea here is that the first part of recognized sequence is delegated to the automaton \(M\) and once this part is recognized, an \(\varepsilon\)-transition takes the automaton \(MN\) to the start state \(q_{0N}\) of \(N\), where the second part of the sequence takes the automaton to an accepting state.

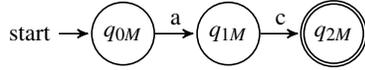
\begin{figure}
\begin{tikzpicture}[->,>=stealth',shorten >=1pt,auto,node distance=1.4cm,
                    semithick,label=foo]

  \node[initial,state]           (0)              {$q_{0M}$};
  \node[state]                   (1) [right of=0] {$q_{1M}$};
  \node[state,accepting]         (2) [right of=1] {$q_{2M}$};

  \path   (0) edge              node  {a} (1)
          (1) edge              node  {c} (2);
\end{tikzpicture}
\caption{A DFA recognizing the sequence ``ac''.}
\label{fig:concatexampleM}
\end{figure}

An example of this is shown in figure \ref{fig:concatexampleMN}, as an automaton constructed from the automata in figure \ref{fig:concatexampleM} and \ref{fig:concatexampleN}.

\begin{figure}
\begin{tikzpicture}[->,>=stealth',shorten >=1pt,auto,node distance=1.4cm,
                    semithick,label=foo]

  \node[initial,state]           (0)              {$q_{0N}$};
  \node[state]                   (1) [right of=0] {$q_{1N}$};
  \node[state,accepting]         (2) [right of=1] {$q_{2N}$};

  \path   (0) edge              node  {d} (1)
          (1) edge              node  {c} (2);
\end{tikzpicture}
\caption{A DFA recognizing the sequence ``dc''.}
\label{fig:concatexampleN}
\end{figure}
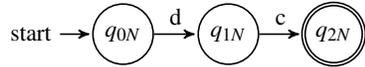

\begin{figure}
\begin{tikzpicture}[->,>=stealth',shorten >=1pt,auto,node distance=1.4cm,
                    semithick,label=foo]

  \node[initial,state]           (0)              {$q_{0M}$};
  \node[state]                   (1) [right of=0] {$q_{1M}$};
  \node[state]                   (2) [right of=1] {$q_{2M}$};
  \node[state]                   (3) [right of=2] {$q_{0N}$};
  \node[state]                   (4) [right of=3] {$q_{1N}$};
  \node[state,accepting]         (5) [right of=4] {$q_{2N}$};

  \path   (0) edge              node  {a} (1)
          (1) edge              node  {c} (2)
          (2) edge              node  {$\varepsilon$} (3)
          (3) edge              node  {d} (4)
          (4) edge              node  {c} (5);
\end{tikzpicture}
\caption{Concatenating automata in figures \ref{fig:concatexampleM} with \ref{fig:concatexampleN} recognizing the sequence ``acdc''.}
\label{fig:concatexampleMN}
\end{figure}
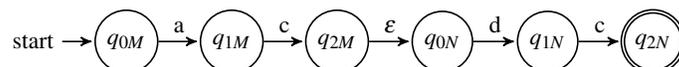

\subsection{Context-free Grammars}
\label{sec:cfg}
Context-free grammars \cite{Autebert97context-freelanguages} (CFG), place less restrictions on their production forms than regular grammars, the left-hand side of every production must contain a single nonterminal symbol.
They can thus be used to describe a larger class of languages than regular grammars. A language described by a CFG is called a context-free language. A typical application of CFGs is to describe programming languages, as those form a subset of the context-free languages. We will also make use of this convenience, when describing domain specific languages (DSLs) later.

\section{Unit Testing in Message Passing Systems}
\label{sec:approach}

Consider that the basic entity for building message-passing systems, usually, is an actor or a component based on the particular systems.
Regardless of the particular system involved, each such message-passing entity, which we refer to generically as a component, can be considered as a state machine. Its execution can be described by the performance of a sequence of atomic events, each of which serve as a transition label in the state machine.
These events can be categorized as input, output and internal events \cite{lynch89}.
\textit{Internal events} such as the scheduling and execution of event handlers occur internally within the component while input and output events, further labelled as \textit{external actions}, correspond to the receipt of incoming events and emission of outgoing events respectively by the component. External events for a component occur at the interface between the component and its environment.

In accordance with the definition of unit testing \cite{IEEESEVocab2010}, we would like to assert assumptions against the behavior of some component under test (CUT) and its interaction with the environment --- the goal being to increase our confidence in the component's implementation.
In the context of unit testing message-passing components, we adopt black-box testing \cite{IEEESEVocab2010} as testing that ignores the internal implementation of the CUT and uses general knowledge of input and output events to verify the behavior of the CUT and white-box testing \cite{IEEESEVocab2010} as testing that uses internal implementation knowledge of the CUT to verify its functionality, for example, by running assertions on its internal state.
To thus verify the behavior of the CUT against a desired protocol, the tester must provide some test specification containing a description of the testable expected behavior, while the test framework, on the other hand, takes on the responsibility of executing an instance of the CUT and verifying its execution against the provided specification.
This raises the question of what the contents of such a specification should be and how it assists the test framework in the verification process at runtime.

\subsubsection{Executions}
\label{sec:executionSequences}
We define an \textit{execution sequence} (or \textit{execution} if this is clear from the context) of a component \textit{c} to be the sequence of external events performed by a single instance of \textit{c} when run in a given environment.
Such an environment comprises \textit{c}, along with other peer components as well as all messages sent between these peers.
In other words, an execution may be considered as a linear arrangement of input and output actions, similar to a string of characters in formal languages, that models the behavior of a component.

\subsubsection{Black-Box Testing as Well-formedness of Executions}
\label{sec:wellformedness}
The act of black box-testing a component becomes equivalent to verifying the correspondence of actual external events of the CUT at runtime against the expected behavior of the CUT.
This means that a tester describes the set of sequences, that are deemed to be correct for a given test case, as the specification, while the framework matches the actual occurrence of events, obtained by running an instance of the CUT, against this specification.
This problem becomes equivalent to the decidability of the well-formedness of a word in the field of formal languages --- that is, whether or not a given word (sequence of symbols) belongs to a language. A task very well suited for finite automata in the case of regular languages as discussed in section \ref{sec:formalgrammar}.

Given that an execution is composed of a finite sequence of events, the set of all such events forms an alphabet, over which a language can be described.
The tester thus specifies a language while the word to be tested for well-formedness is the execution sequence observed by the test framework at runtime by listening on events incoming to and outgoing from the component.
The idea of black box-unit testing thus can be summarized as the act of writing a specification describing a language \(\mathrm{L}\) of all acceptable executions for that particular test case, while the test framework verifies the runtime execution of the CUT against \(\mathrm{L}\).

\subsubsection{Formal Grammar for Specifying Executions}
Our approach uses a mechanism for specifying execution sequences as regular languages, allowing the utilization of regular expression techniques for writing tests.
As a result, a test specification may be converted into a finite automaton by the testing framework such that the test passes if the automaton ends in an accepting state.
This allows for the runtime execution of the CUT while observed events are being matched by the automaton simultaneously.
An added advantage of this is that it enables the tester to use likely familiar concepts from regular expressions when writing tests.

\subsection{A Test Specification for Executions}
With the aim of producing a framework for performing automated and repeatable testing, we desire the test specification to be unambiguous as well as machine executable.
For this reason, a test specification can be considered as a program, written in some language, not to be confused with the language of execution sequences it describes.
The test framework implementation then acts as a machine executing this program, responsible for verifying the correspondence of the CUT to the specification.

An advantage of using a specialized language is that it enables the specification of instructions for additionally controlling the behavior of the test framework, resulting in more concise and flexible test specifications.
For brevity, we assume that environment setup activities such as the actual creation of CUT instances and other participating components and their dependencies are handled prior to program execution.

The following sections introduce the theoretical framework using syntatic constructs that generate a test specification.
The formation rule used for illustration is a context-free grammar that describes a block-structured language for specifications. The CFG allows us to succinctly describe most of the constructs, independent of any particular message-passing system.
We show how these constructs can be mapped to automata and where applicable, how they correspond to their regular expression variant.
The discussions also include how they can be used to write tests for the vast majority of desired behaviors in both deterministic and non-deterministic environments.

\subsection{Regular Operations on Execution Sets}
In this section, we illustrate the variants of regular expression operations offered by our framework for describing events comprising execution sequences and building larger executions from smaller ones.
In particular, we illustrate the union and concatenation of the languages described by a specification while the Kleene closure operation is explained in Section \ref{sec:closure} when the concept of blocks have been introduced.

\subsubsection{Concatentation of Execution Sets}
\begin{lstlisting}[label=lst:concat, float=t!, caption={Concatenation of executions}]
Body -> expect Event$^+$
Event -> $e$
\end{lstlisting}

\begin{figure}
\begin{tikzpicture}[->,>=stealth',shorten >=1pt,auto,node distance=1.4cm,
                    semithick,label=foo]

  \node[initial,state]           (0)              {$q_0$};
  \node[state,accepting]         (1) [right of=0] {$q_1$};

  \path   (0) edge              node  {$e$} (1);
\end{tikzpicture}
\caption{An automaton recognizing a single event $e$.}
\label{fig:dfaSingleEvent}
\end{figure}
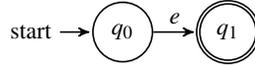

Listing \ref{lst:concat} shows the production for concatenating a sequence of single events using the \lstinline!expect! keyword. The symbol \(^+\) means one or more occurrences. Statements that appear on the right hand side of the \emph{Body} non-terminal describe a unique language over the set of event alphabet, that is, they describe a unique set of execution sequences.
As with the symbols of regular expressions, a single event \(e\) matches itself and the language \(L(e)\) described by \(e\) consists only of itself --- that is \(L(e) = \{e\}\).
In terms of the automaton created, at runtime, the statement ``\lstinline!expect ev_!'' would cause the automaton to move from the start to next and final state only when \(e\) has occurred --- thus matching \(e\). Any other event observed at this state leads to a failed test case.
Figure \ref{fig:dfaSingleEvent} shows a deterministic finite automaton DFA \(M_e\) recognizing a single event \(e\).

For concatenation of sets of executions, we also use the definition of concatenation of regular languages as defined in \ref{sec:regulargrammar}.
The \emph{expect} keyword allows us to specify a list of execution sets to be concatenated.
As an example, the statement \lstinline!expect e1e2e3! describes the language \(L = \{e_1e_2e_3\}\) formed by concatenating the three languages \(L(e_1) = \{e_1\}\), \(L(e_2) = \{e_2\}\) and \(L(e_3) = \{e_3\}\).

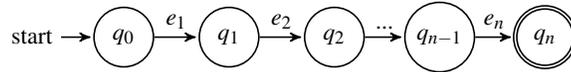
\begin{figure}
\begin{tikzpicture}[->,>=stealth',shorten >=1pt,auto,node distance=1.4cm,
                    semithick,label=foo]

  \node[initial,state]           (0)              {$q_0$};
  \node[state]                   (1) [right of=0] {$q_1$};
  \node[state]                   (2) [right of=1] {$q_2$};
  \node[state]                   (3) [right of=2] {$q_{n-1}$};
  \node[state,accepting]         (4) [right of=3] {$q_n$};

  \path   (0) edge              node  {$e_1$} (1)
          (1) edge              node  {$e_2$} (2)
          (2) edge              node  {...}   (3)
          (3) edge              node  {$e_n$} (4);
\end{tikzpicture}
\caption{Automaton for recognizing the concatenation of events.}
\label{fig:concat}
\end{figure}

Generally, given a statement \(S = \) ``\lstinline!expect e1 e2 e3 ... e_n!'', where each \(a_i\) describes a unique language \(L(a_i)\), a DFA \(M_S\) recognizing \(L(S)\) is created by concatenating each automaton \(M_{ai}\) that recognizes \(L(a_i)\) sequentially. Figure \ref{fig:concat} shows a DFA \(M_S\) constructed in this manner.

\subsubsection{Union of Executions Sets}
\label{sec:union}
\begin{lstlisting}[label=lst:union, float=t!, caption={Union of executions}]
Body -> either Body$^+$ or Body$^+$ end
\end{lstlisting}

Listing \ref{lst:union} extends the grammar in Listing \ref{lst:concat} with constructs for creating the union of execution sets via the \textbf{either}-\textbf{or} conditional statement.
As each \emph{Body} nonterminal describes a unique language, the conditional statement contains two independent langauges from the \emph{either} and \emph{or} branches which are combined via the union operation on regular languages to form the described language of the conditional statement.

Conditional statements allow the tester to describe possible paths within the state space of the component, exactly one of which would be traversed depending on the observed event at runtime.
These are convenient in situations where there are several, alternative and possibly equivalent paths outgoing from a certain state of the CUT.
In some cases, it may be inconvenient or difficult to reproduce a test environment that consistently guides the framework through any one of the alternate paths, while in other cases, the paths may be supplied to increase the robustness of the test case --- the test case may be designed so that a random path is chosen at each iteration.

With the \emph{Body} nonterminal, any statements are allowed inside the \emph{either} and \emph{or} branch of a conditional statement, including other conditional statements.

Given a conditional statement \(S\) with statements \(A\) and \(B\) as specified by its \emph{either} and \emph{or} branches respectively, the language \(L(S)\) described by \(S\) is defined to be the union of the sets described by both branches, that is \(L(S) = L(A) \cup L(B) \).
Consequently, an automaton \(M_S\) recognizing \(L(S)\) would be equivalent to the automaton \(M_{A \cup B}\) that recognizes \(L(A) \cup L(B)\).

We construct an NFA for this purpose by combining sub-automata \(M_A\) and \(M_B\), for \(L(A)\) and \(L(B)\) respectively, alongside each other as described in \cite{hopcroftullman79}.
All states and transitions of the sub-automata remain throughout the construction.
A new start state \(q_{A \cup B}\) of \(M_S\) is created by merging the start states \(q_A\) and \(q_B\) of \(M_A\) and \(M_B\) respectively.
This new state will have the same outgoing transitions as the combined start states, allowing the NFA to reach the final states of either sub-automaton when verifying events at runtime.
The set of final states of the NFA is the union of the final states of both sub-automata since the NFA accepts any execution that is accepted by either sub-automaton.

\begin{figure}
\begin{tikzpicture}[->,>=stealth',shorten >=1pt,auto,node distance=1.4cm,
                    semithick,label=foo]

  \node[initial,state]           (0)              {$q_{A \cup B}$};
  \node[state]                   (1) [above of=0] {$q_{A0}$};
  \node[state]                   (2) [right of=1] {$q_{A1}$};
  \node[state,accepting]         (3) [right of=2] {$q_{A2}$};
  \node[state]                   (4) [below of=0] {$q_{B0}$};
  \node[state]                   (5) [right of=4] {$q_{B1}$};
  \node[state,accepting]         (6) [right of=5] {$q_{B2}$};

  \path   (0) edge              node  {$e_1$} (2)
              edge              node  {$e_1$} (5)
          (1) edge              node  {$e_1$} (2)
          (2) edge              node  {$e_2$} (3)
          (4) edge              node  {$e_1$} (5)
          (5) edge              node  {$e_3$} (6);
\end{tikzpicture}
\caption{NFA for conditional statement: ``\lstinline!either expect e1 e2 or expect e1 e3 end!''.}
\label{fig:conditional}
\end{figure}
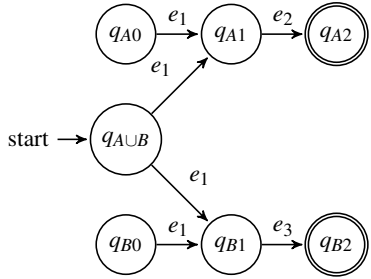

As an example, consider the statement \(S = \) ``\lstinline!either expect e1 e2 or expect e1 e3 end!''.
Figure \ref{fig:conditional} shows the constructed NFA \(M_S\) for this statement. The start state \(q_{A \cup B}\) contains the same transitions outgoing from the start states \(q_{A0}\) and \(q_{B0}\) of the sub-automata while the final states \(q_{A2}\) and \(q_{B2} \) of the NFA are those of the sub-automata.
The constructed NFA can be subsequently be converted to a DFA or simulated at runtime using a set of states that keep track of the possible states of the automaton.

\subsection{Blocks and Executions}
As mentioned, our illustrative specification language is a block-structured language. This means that the language allows for the creation of blocks as well as nested blocks, enabling the tester to group a sub-sequence of events into a single unit.

\begin{lstlisting}[label=lst:exec, float=t!, caption={Production for generating an execution.}]
Exec -> repeat num? Hdr$^\ast$ body Body$^\ast$ end
\end{lstlisting}

A block is the topmost construct of a test specification. As shown in Listing \ref{lst:exec}, it generates an execution which may be made up of smaller sub-executions using nested blocks. Here, \(n\) is a positive number while \(?\) means that it is optional.
Every specified event belongs to a single block.
As with a lot of block-structured programming languages, the benefits of this are manifold --- a scope can be invoked throughout a block, allowing declarations of constraints and requirements on events that only affect the block's sub-sequence at test runtime. For example, a tester may want to specify that some particular event must be present within a block's sequence.
For example, consider the execution \(e_1e_2e_3e_4\), the tester may declare that the sub-sequence \(e_2e_3\) delimit a nested block so that \(e_1\) and \(e_4\) continue to be associated with the outer block.
Now within the newly declared block, the tester may want to declare a requirement that some event \(e_0\) must occur at some point within the block --- hence the actual described sequence is \(e_1e_0e_2e_3e_4|e_1e_2e_0e_3e_4|e_1e_2e_3e_0e_4\). This is further explained in Section \ref{sec:blockrequirements}

The production in Listing \ref{lst:exec} splits a block into a \textit{header} and a \textit{body} identified by the \lstinline!Hdr! and \lstinline!Body! nonterminals.
The \lstinline!Hdr! non-terminal generates constraints and requirements on the events observed within the entire block as well as nested blocks while the \lstinline!Body! as mentioned generates executions belonging to the block.

The language of an entire block is the set of execution sequences described by applying the specified constraints and requirements declared in the header, to the execution sequences described by the body.
The entire block is declared within keywords \emph{repeat} and \emph{end} with an optional positive integer \emph{n}.
This is used to generate a block that consecutively matches a sub-sequence at runtime, the specified number of times or in the case without any number, a block that matches the Kleene closure on the defined language.

\subsubsection{Matching Repeated Executions}
\label{sec:repeat}
The block construct, when declared with a positive integer \textit{n},  instructs the framework at runtime to match an sub-sequence against its described execution sequence \textit{n} times.
This mechanism facilitates concise test specifications, allowing the tester specify a repeating sequence of events only once.
For example, if the tester expects an execution of the form ``\lstinline!abcabcabc...!'' where \(a, b, c\) are events, the sequence may only be declared once as ``\lstinline!repeat num body expect a b c!'' where \(n\) is the number of expected occurences of the subsequence ``\(abc\)''.

Suppose that a DFA \(M_s\) recognizes the language of a block \(s\) without a \textit{repeat} operation invoked. What we know about the actual language \(S\) described by the block, with the \textit{repeat} operation invoked, is that it is a language formed by concatenating \(s\) with itself \textit{n} times.
An execution in \(S\) is formed by taking any \textit{n} executions in \(s\) and concatenating them.
As a result, a DFA \(M_S\) recognizing \(L(S)\) can be constructed using \(n\) copies of the automaton \(M_s\) by concatenating them together.

Consider as an example, the statement \(S = \) ``\lstinline!repeat 2 body expect e1 e2 end!''.
\(L(S) = \{e_1 e_2 e_1 e_2\}\). The language described by the block \(B\) before the \emph{repeat} operation is invoked is \(L(B) = \{e_1 e_2\}\).
However, the sequence is expected twice, resulting in the final language described by the automaton \(M_B M_B\) where \(M_B\) is an automaton recognizing \(L(B)\).

\subsubsection{Matching the Kleene Closure of an Execution Set}
\label{sec:closure}
The \textit{repeat} statement, when declared without a positive integer, instructs the framework to match the Kleene closure on the described language. The framework matches zero or more occurrences of observed executions that belong to the described language.
Analogous to the Kleene closure on a language, we denote such a statement \(S^*\) and define its language \(L(S^*)\) as the set of executions that can be formed by concatenating any number of executions described by \(S\), the associated block.
For example, given a block \(B\) describing the set of executions \(L(B) = \{e_1,e_2\}\), \(L(B^*)\) describes the set consisting of all executions containing only the actions \(e_1\) and \(e_2\) --- i.e. \(\{e_1, e_2, e_1 e_2, e_2 e_2, ...\}\).
Since the closure of a language matches zero or more occurrences, the language \(L(S^*)\) always includes an empty execution (containing no events) regardless of the execution described by \(S\).

The addition of the closure operation into our test specification neccessarily introduces nondeterminism when verifying the occurrence of events at runtime using an automaton.
Consider the language \(L(S^*)\) of some block \(S\).
An equivalent automaton \(M_{S^*}\) at runtime, that matches some initial execution \(E_0 \in L(S)\) must transition to a next state that implies the current state of the CUT, without having access to future events which might be yet to occur.
In such a case, the automaton must correctly guess between two options ---  a transition to a final state signalling that it is done matching executions in \(L(S)\), or a transition to a next state that expects to match another execution \(E_1 \in L(S) \).

In accordance to the construction of a finite automaton for the Kleene closure of a language \cite{hopcroftullman79}, we build a nondeterministic finite automaton with epsilon transitions (\(\varepsilon\) -NFA) as \(M_{S^*}\) to recognize the closure on an execution set described by some block \(S\).
To do this, we start with the automaton \(M_S\) that recognizes \(L(S)\)
and transform \(M_S\) into \(M_{S^*}\) by introducing two types of \(\varepsilon\)-transitions corresponding to the choices to be made by the automaton.
We form an \(\varepsilon\)-transition from the start state \(q_0\) of \(M_S\) to every final state of \(M_S\) allowing the automaton to go directly to the final state when it guesses that all executions have been verified.
We also form \(\varepsilon\)-transitions from each final state \(q_n\) back to \(q_0\), allowing any number of execution sequences described by \(S\) to be verified by the automaton.

\begin{figure}
\begin{tikzpicture}[->,>=stealth',shorten >=1pt,auto,node distance=1.8cm,
                    semithick,label=foo]

  \node[initial,state]           (0)              {$q_0$};
  \node[state]                   (1) [right of=0] {$q_1$};
  \node[state,accepting]         (2) [right of=1] {$q_2$};

  \path   (0) edge              node  {$e_1$} (1)
              edge [bend left]  node  {$\varepsilon$} (2)
          (1) edge              node  {$e_2$} (2)
          (2) edge [bend left]  node  {$\varepsilon$} (0);
\end{tikzpicture}
\caption{$\varepsilon$-NFA for ``\lstinline!repeat body expect e1 e2 end!''.}
\label{fig:closure}
\end{figure}
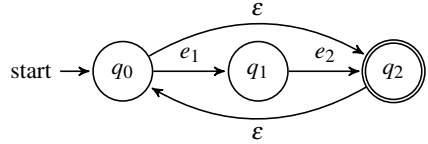

As an example, figure \ref{fig:closure} shows an \(\varepsilon\)-NFA \(M_{S^*}\) recognizing the language described by the specification \lstinline!repeat body expect e1 e2 end!. Here, the framework transitions directly from the \(q_0\) to \(q_2\) if no actions are observed while the the path from \(q_0\) to \(q_2\) via \(q_1\) is traversed \(n\) times where \(n\) is the number of consecutive \(e_1 e_2\) executions that occur at test runtime.

Such an automaton could be directly converted to a DFA using the subset construction or simulated directly by keeping a set of states that the automaton could possibly be in as facilitated by the eclosure mechanism \cite{hopcroftullman79}.

\subsection{Ordering an Execution}
\label{sec:orderingActions}
Accurately predicting the order of events that make up an execution becomes problematic in message-passing systems with multiple components, due to the inherent asynchrony involved.
This is expecially problematic in asynchronous distributed environments where there are no upper bounds on computation and message transmission time \cite{raynal96, lamport78}.
However, this problem becomes manageable in the scope of unit testing since the only matched events are local to a single component in the system.
Inevitably, the class of verifiable scenarios by the framework are limited to those of local properties of the CUT.
For example, scenarios verifying global properties of algorithms, which likely involve asserting properties across several components, are not specifiable. Nonetheless, it sufficiently serves the purpose of unit testing.

In the scope of unit testing of a single component in a message passing system, we consider inherent problems of nondeterminism such as the unpredictable scheduling of components, lack of upper bounds on computation steps and message delays as it pertains to the events observed locally at the CUT's interface.
For example, a set $M$ of outgoing requests from the CUT to a set of peer components may expect all incoming responses in a set $N$. The order in which the requests are sent or responses arrive at the CUT may not be accurately predictable when writing the tests.
Hence a need to explicitly specify a sequence of unordered actions for nondeterministic events in an execution.

\begin{lstlisting}[label=lst:unordered, float=t!, caption={Generating an unordered execution.}]
Event -> $e$ | unordered $e^+$  end
\end{lstlisting}

Listing \ref{lst:unordered} generates an unordered sequence of events which are specified between the \textit{unordered} and a matching \textit{end} keyword.
Since the order of the events do not matter, the language described by the statement consists of all permutations of the originally specified sequence.

As an example, the following statement ``\lstinline!expect e1 e2  unordered e3 e4 end e5''! instructs the framework to fail the test case at runtime on observing, say, the execution ``\lstinline!e2e1e3e4e5!'' but not on executions ``\lstinline!e1e2e3e4e5!'' and ``\lstinline!e1e2e4e3e5!''.

\subsubsection{Constructing Automata for Unordered Executions}
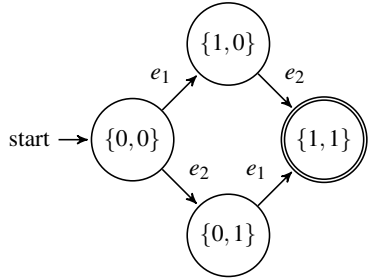
\begin{figure}
\begin{tikzpicture}[->,>=stealth',shorten >=1pt,auto,node distance=1.8cm,
                    semithick,label=foo]

  \node[initial,state]           (0)              {$\{0,0\}$};
  \node[state]                   (1) [above right of=0] {$\{1,0\}$};
  \node[state]                   (2) [below right of=0] {$\{0,1\}$};
  \node[state, accepting]        (3) [below right of=1] {$\{1,1\}$};

  \path   (0) edge              node  {$e_1$} (1)
              edge              node  {$e_2$} (2)
          (1) edge              node  {$e_2$} (3)
          (2) edge              node  {$e_1$} (3);
\end{tikzpicture}
\caption{DFA for ``\lstinline!unordered e1 e2 end!''; key $\{e_1,e_2\}$.}
\label{fig:dfaUnordered}
\end{figure}

Consider the statement \(S = \) ``\lstinline!expect unordered e1 e2 end!'' describing the language \(L(S) = \{e_1 e_2, e_2 e_1\}\).
Figure \ref{fig:dfaUnordered} shows an automaton \(M_S\) recognizing an equivalent language.
Generally, the statement ``\lstinline!unordered e1 e2 ... e_n end!'' for \(n\) actions \(e_1\) to \(e_n\) describes a language containing \(n!\) executions --- each, a permutation of the originally specified sequence.
As a result, a DFA \(M\) describing an equivalent language must accept exactly all \(n!\) possible sequences.

One way to build \(M\) is to construct a path for each possible sequence from the start to the end states of the automaton.
Each path contains exactly \(n\) transitions and \(n + 1\) states and each state within a path is used to remember which events have been matched and which are pending.
Such a state can be thought of as being associated with a bit string of length \(n\) representing the set of specified events \(e_1\) to \(e_n\) such that the \(ith\) bit is 1 if the event \(e_i\) has been matched in the execution and 0 otherwise.

Consider the start state \(q_0\) of \(M\). No events have been matched at this state so all bits are set to 0 - that is, the bit string associated with this state is ``\(000...0\)''.
Now suppose at runtime that the event \(e_n\) occurs first. Then the automaton transitions to the next state \(q_\delta\) associated with the bit string ``\(000...1\)'' with only the \(nth\) bit set to 1.
A transition from \(q_\delta\) on action \(e_1\) leads to the next state \(q_\gamma = \) ``100...1'', and so on with the final state of the automaton \(q_\phi = \) ``111...1'' signalling that all specified events have been matched.

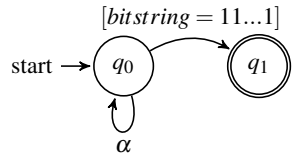
\begin{figure}
\begin{tikzpicture}[->,>=stealth',shorten >=1pt,auto,node distance=1.8cm,
                    semithick,label=foo]

  \node[initial,state]           (0)              {$q_0$};
  \node[state,accepting]         (1) [right of=0] {$q_1$};

  \path   (0) edge[bend left]              node  {$[bit string = 11...1]$} (1)
          (0) edge[loop below]             node  {$\alpha$} (1);
\end{tikzpicture}
\caption{EFSM for unordered executions.}
\label{fig:efsmUnordered}
\end{figure}

This mechanism generates a total of \(n!\) states where \(n\) is the number of specified events, making it impractical for even modest values of \(n\).
In implementation however, the need for extra states is easily avoided since unlike with DFAs, the task of remembering the matched actions can be accomplished programmatically by the framework.
Figure \ref{fig:efsmUnordered} shows a more practical scheme using an EFSM - an extended finite state machine \cite{alagar11}. The self-transition labelled \(\alpha\) represents the set of events specified by the \emph{unordered} statement.
The automaton uses exactly 2 states regardless of the size of \textit{n} and enables the transition from the start to the end state only when the trigger condition, that all bits in the string have been set to 1, has been met.

\subsection{Specifying Constraints and Requirements on Executions}
\label{sec:constraintsOnEvents}

Having a programmable test specification allows the possibility of placing flexible restrictions on the events observed within the execution.
Listing \ref{lst:header} shows the right hand of the \textit{header} nonterminal \emph{Hdr} that is used to generate such restrictions in a test specification.
The semantics of these constructs assume that, unlike events specified using the \textit{expect} keyword, it is not possible to predict an exact instance in the test execution where the constraints must be satisfied.
As a result, they apply to an entire block and any nested blocks in the specification. For this reason, they can only be specified within the \emph{header} of a block.

\begin{lstlisting}[label=lst:header, float=t!, caption={Generating restrictions on observed executions}]
Hdr  -> allow $e^+$ | disallow $e^+$ | drop $e^+$ | blockExpect $e^+$
\end{lstlisting}

\subsubsection{Constraints on Blocks}
Within a block, some events may be \textit{disallowed} by the tester. The presence of such events at any point when the framework executes a statement in the block is undesirable so that it is unnecessary to continue running the test case - the test case should terminate immediately with a failure.
In some other cases, the occurence of certain events in an execution is not neccessary to validate the test case. In fact, such events may not even be observed in multiple executions of the same test case. In other words, these events are not required for a successful test case but are \textit{allowed} if they occur.
Finally, the messages associated with some events may be \textit{dropped} on occurrence. This means that such messages, if outgoing from the CUT, should not be forwared to recipients and if incoming, should not be delivered to the CUT.
Such a mechanism is particularly useful when writing test logic that verify edge cases and error conditions by enabling the tester guide the CUT into the vulnerable state. However, it does require knowledge of events that should or should not be handled by the CUT in order to enter such a state be available to the tester.

As is typical of block-structured languages, constraints are only valid within the scope of the block where it was declared as well as its nested blocks.
However, a constraint \(C_1\) on an event \(e\) in block \(B_1\) can be \textit{shadowed} by redeclaring a new constraint \(C_2\) on \(e\) in nested block \(B_2\) of \(B_1\).
This means that \(C_2\) is valid if \(e\) is observed within \(B_2\) and \(C_1\) is valid if \(e\) is observed within \(B_1\) but not \(B_2\).
Conflicts occur when constraints on the same events are declared within the same block. In such cases, only the last constraint declared is enforced.

As much as these statements within block headers describe the execution sequence, they also control the behavior of the framework.
For example the \textit{disallow} statement may be interpreted as an instruction to the framework to fail the testcase if any of the specified events appear within the sequence while \textit{allow} and \textit{drop} instruct the implementation whether or not to forward the specified events if they do appear in the sequence.
In other words, from the perspective of building an automaton that recognizes an execution, actions specified by these constructs do not neccessarily drive the automaton closer to a final state.

\begin{figure}
\center
\begin{tikzpicture}[->,>=stealth',shorten >=1pt,auto,node distance=1.4cm,
                    semithick,label=foo]

  \node[initial,state]           (0)              {$q_0$};
  \node[state]                   (1) [right of=0] {$q_1$};
  \node[state, accepting]        (2) [right of=1] {$q_2$};

  \path   (0) edge              node  {$e_1$} (1)
          (1) edge              node  {$e_2$} (2);
\end{tikzpicture}
\caption{Automaton recognizing ``\lstinline!repeat 1 body expect e1 e2 end!''.}
\label{fig:dfaExpecte1e2}
\end{figure}
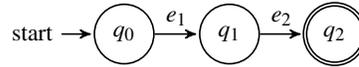

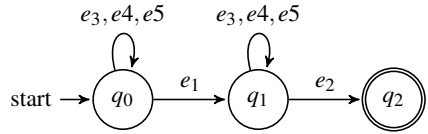
\begin{figure}
\center
\begin{tikzpicture}[->,>=stealth',shorten >=1pt,auto,node distance=1.8cm,
                    semithick,label=foo]

  \node[initial,state]           (0)              {$q_0$};
  \node[state]                   (1) [right of=0] {$q_1$};
  \node[state, accepting]        (2) [right of=1] {$q_2$};

  \path   (0) edge              node  {$e_1$} (1)
              edge [loop above] node  {$e_3, e4, e5$} (0)
          (1) edge              node  {$e_2$} (2)
              edge [loop above] node  {$e_3, e4, e5$} (1);
\end{tikzpicture}
\caption{Extending Figure \ref{fig:dfaExpecte1e2} with headers ``\lstinline!allow e3 e4 drop e5!''.}
\label{fig:dfaAllowDrop}
\end{figure}

Consider the specification \(S = \) ``\lstinline!repeat 1 body expect e1 e2 end!'', an equivalent automaton \(M_S\) shown in Figure \ref{fig:dfaExpecte1e2} and the specification \linebreak \(C = \) ``\lstinline!repeat 1 allow e3 e4 drop e5 body expect e1 e2 end!'' having an equivalent block body to \(S\).
To build the automaton \(M_C\) shown in Figure \ref{fig:dfaAllowDrop} recognizing \(L(C)\), we incorporate \textit{allow} and \textit{drop} statements by adding a self transition on the set of specified events, to each state of the automaton for the entire block body including nested blocks \((q_0, q_1)\).
In the case of \(M_C\), this implies that states recognizing events \(e_1\) and \(e_2\) become annotated a self transition on events \((e_3, e_4, e_5)\) as specified by the block header.
No distinction is made between \textit{allow} and \textit{drop}  statements in the automaton, the actual difference is in the behavior of the framework implementation (whether or not it forwards the event).

Since events declared with the \textit{disallow} keyword causes the testcase to fail on occurrence, an equivalent event from the automaton's perspective would be to include a transition leading to an error state on the specified actions --- such transitions are implicit as mentioned in section \ref{sec:dfa}. Just as with \textit{allow} and \textit{disallow} statements, such a transition would be included on each state within the block.

\subsubsection{Requirements on Blocks}
\label{sec:blockrequirements}
Events preceeded by the \emph{blockExpect} keyword as shown in Listing \ref{lst:header} are required the occur at runtime at some point within the block.
This places a requirement on the framework to successfully exit the block not only when all statements have been executed but also that all such required events have been verified.
For example, given the specification ``\lstinline!repeat blockExpect e0 body expect e1 e2 end!. The test framework does not exit successfully until events \(e_1\) and \(e_2\) have been matched in order but also the event \(e_0\) has been observed as well.
Hence, the actual language of this specification is \(\{e_0e_1e_2,\enskip e_1e_0e_2, \enskip e_1e_2e_0\}\) as \(e_0\) may occur at any position in the sequence while the relative positions of all other events are respected.
Since such events are not specified in any order relative to other events within the block, this construct is able to describe a wider range of non-deterministic scenarios than \textit{unordered} statements.

To incorporate the \emph{blockExpect} statement into an automaton, note that any sequence described by this statement must be interleaved with the sequences described by the rest of the block.
Since the statement itself describes its own language (all permuatations of the specified execution), we can distinguish this language from that described by the rest of the block.

We illustrate the construction process with an example using the specification \linebreak \(S = \) ``\lstinline!repeat 1 blockExpect e3 e4 body expect e1 e2 end!''.
The block header declaration \(H = \) ``\lstinline!blockExpect e3 e4!'' describes the language \(L(H) = \{e_3 e_4, e_4 e_3\}\), while the body \(B\) describes the language \(L(B) = \{e_1 e_2\}\).
Let \(h\) be the number of events specified by \(H\) and \(b\) the number of events specified by \(B\). In this case, \(h = b = 2\).
The semantics of the \textit{blockExpect} produces the language \(L(S) = \{s \enskip | \enskip s  \) is a permutation of  \(``e_1 e_2 e_3 e_4" \land e_1 \) appears before  \(e_2\}\).
An automaton \(M\) describing \(L(S)\) must thus recognize only sequences containing exactly the events \(e_1\) through \(e_4\) such that the order imposed by the entire block is preserved.

We start with the automaton \(M_B\) recognizing \(L(B)\). Each state in \(M_B\) remembers the next expected expected as specified in the block body \(B\).
To incorporate \(H\), note that for each state in \(M_B\), \(M\) must remember what actions from \(H\) have already occurred.
This can be done using a bit string of length \(h\) where the \(i^{th}\) bit represents whether or not the \(i^{th}\) event specified by \(H\) has occurred.
Thus each state in \(M\) is annotated with a bit string, as well as the next expected action in \(B\), leading to \(2^h\) states for each event \(e\) in \(B\) as each state for \(e\) remembers a unique possible configuration of the bit string.

To define the transition function \(\delta\) for the automaton M, consider each state as a pair \((e, s)\) where \(e\) is the next expected event in \(B\) and \(s\) is the bit string associated with that state.
Valid transitions from a state \(q\) are only available for an unverified event \(e_n\) such that \(e_n \in I_B\) or \(e_n \in I_H\) - where \(I_B\) and \(I_H\) are the sets of events declared by the body \(B\) and header \(H\) respectively.
For a given state \(q = \{e_i, s\}\), If \(e_n \in I_B\), then the next state is the state \(q_\gamma = \{e_{i+1}, s\}\) else if \(e_n \in I_H\) such that \(e_n\) equals the \(k^{th}\) event specified by \(H\), then the next state is \(q_\gamma = \{e_i, s_\gamma\}\) where \(s_\gamma\) is a copy of \(s\) with the \(k^{th}\) bit set to 1.
The final state \(q_\phi = \{\emptyset, s_\phi\}\) has all bits in \(s_\phi\) set to 1 and \(e\) empty since no events are expected.

\begin{figure}
\begin{tikzpicture}[->,>=stealth',shorten >=1pt,auto,node distance=1.8cm,
                    semithick,label=foo]

  \node[initial,state]           (0)                    {$\{e_1,0\}$};
  \node[state]                   (1) [above right of=0] {$\{e_2,0\}$};
  \node[state]                   (2) [below right of=0] {$\{e_1,1\}$};
  \node[state]                   (3) [right of=2]       {$\{e_2,1\}$};
  \node[state]                   (4) [right of=1]       {$\{\emptyset,0\}$};
  \node[state, accepting]        (5) [below right of=4] {$\{\emptyset,1\}$};

  \path   (0) edge              node  {$e_1$} (1)
              edge              node  {$e_3$} (2)
          (1) edge              node  {$e_3$} (3)
              edge              node  {$e_2$} (4)
          (2) edge              node  {$e_1$} (3)
          (3) edge              node  {$e_2$} (5)
          (4) edge              node  {$e_3$} (5);
\end{tikzpicture}
\caption{DFA for ``\lstinline!repeat 1 blockExpect e3 body expect e1 e2 end!''; key \{nextaction in body, bitstring of blockExpect\}.}
\label{fig:dfaBlockExpect}
\end{figure}
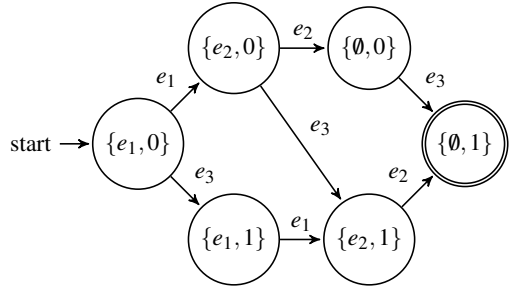

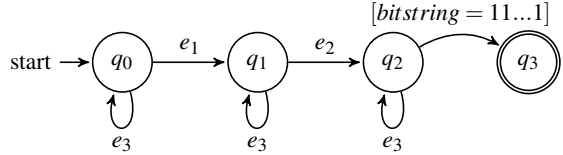
\begin{figure}
\begin{tikzpicture}[->,>=stealth',shorten >=1pt,auto,node distance=1.8cm,
                    semithick,label=foo]

  \node[initial,state]           (0)                    {$q_0$};
  \node[state]                   (1) [right of=0]       {$q_1$};
  \node[state]                   (2) [right of=1]       {$q_2$};
  \node[state,accepting]         (3) [right of=2]       {$q_3$};

  \path   (0) edge              node  {$e_1$} (1)
              edge [loop below] node  {$e_3$} (1)
          (1) edge              node  {$e_2$} (2)
              edge [loop below] node  {$e_3$} (2)
          (2) edge [bend left]  node  {$[bitstring = 11...1]$} (3)
              edge [loop below] node  {$e_3$} (3);
\end{tikzpicture}
\caption{EFSM for Figure \ref{fig:dfaBlockExpect}.}
\label{fig:efsmBlockExpect}
\end{figure}

Figure \ref{fig:dfaBlockExpect} shows an automaton constructed using this technique for the specification ``\lstinline!repeat 1 blockExpect e3 body expect e1 e2 end!''
This technique creates a \((b+1)2^h\) states for the constructed automaton. Just as with the automaton constructed in Section \ref{sec:orderingActions}, an implementation similar to an extended finite state machine can accomplish the task using \(b + 2\) states.
Such an automaton uses \(b + 1\) states to represent the body of the specification with the logic for remembering previously verified events in \(H\) delegated to the framework as opposed to the automaton itself. As a result, we include an additional \textit{sink} state before the final state that transitions to the final state only when all actions in \(H\) have been verified.
Figure \ref{fig:efsmBlockExpect} shows such an EFSM with the transition from the sink state \(q_2\) to the final state \(q_3\) enabled only when all required events specified by the \emph{blockExpect} have been verified, in this case \(e_3\).

\subsection{Request-Response Patterns}
\label{sec:reqres}
In some cases a tester may expect a set of matching incoming messages (responses) in response to a set of outgoing messages (requests).
A variation of which can be accomplished using a specification similar to ``\lstinline!expect unordered $rq_1$ $rq_2$ ... $rq_n$ $rs_1$ $rs_2$ ... $rs_n$ end!'' where \(rs_i\) is the matching response for request \(rq_i\).
The requests and responses are expected in any order, accounting for the non-determinism in response order.
In typical request-response scenarios, there exists a way of uniquely  identifying the response to a request, for example using a unique id \(i\) shared by both \(rq_i\) and \(rs_i\).
However, the example specification assumes knowledge of this id and this may not always be predictable at test specification time.
It is not unusual for such ids to be randomly generated by the CUT or some other artefact in the system.
Thus we need a mechanism for matching responses for requests with randomly generated ids.

We note that the language described by this problem is not regular as it requires some context in order to determine whether or not a response matches some previous request.
Since we do not have a unique identity of the requests at specification time to begin with, we need some way to set a checkpoint when the automaton starts expecting requests and subsequent responses.
The tester describes some property of the requests, like the class name in for example Java, as well as a mapping function \(\alpha: E \times E \to \mathbb{B}\) where \(\mathbb{B} = \{true, false\}\) for that request, that takes a request and a response events and returns true if the provided request matches the response.
Such a function may simply compare the ids of the messages associated with both events.
The framework implementation on reaching the defined checkpoint, keeps a set of observed requests and for each response received, uses the matching function to determine if it matches some requests, advancing the automaton as needed.

\subsection{White Box Testing and Dependencies}
\subsubsection{Testing with external dependencies}
\label{sec:trigger}

Normally, the CUT communicates with other peer components at runtime in its deployed environment in order to complete its computations. These peers thus become dependencies when testing and must be present or simulated in order to correctly exercise the CUT.
There are several reasons for decoupling the CUT from its dependencies  when unit testing. Tests themselves becomes less complex to write when the logic does not additionally need to coordinate peer components in order to reproduce the necessary conditions for the test.
Also, testing with actual dependencies may also cause computations of other components to cross the boundaries of the unit test, for example, when a peer happens to be a component located on a different network and network messages are not desirable in the tests.
More important is the certainty that the source of a test failure does not originate external to the CUT --- a failure in some peer component may cause the test case to inaccurately verify the CUT.

When such situations are undesirable, a solution is usually to replace these dependencies with mocks \cite{mackinnon00} --- deterministic and usually less complex versions of a dependency offering the minimal features that facilitate the test execution.
This requires no new constructs with respect to our proposed framework as no distinction is made between a mock and a real version of a component in the system.
Both parties are simply components interacting with the CUT, allowing the tester to decide to mock on a per-component basis. Note that it may be simpler to forego the mock and use the actual dependency in some cases after all.

\begin{lstlisting}[label=lst:trigger, float=t!, caption={Interactively producing events}]
Body -> trigger $e^+$
\end{lstlisting}

The second option offered by the framework is to send events interactively into the test environment. This is done using the the \emph{trigger} construct within the body of a block as shown in Listing \ref{lst:trigger}. It can thus be used to mimic some peer component's participation in the test scenario by sending messages to the CUT instance at runtime as if they were sent by the peer itself. 
This option produces more flexible but potentially verbose test specifications diffused with operations.
A \emph{trigger} statement prefixes a list of events, instructing the test framework to send the messages associated with those events to the CUT. In fact, this operation needs not be limited to CUT events. Specified events may be destined for any peer component in the test environment as long as the framework is able to provide some mechanism for identifying the recipient. Thus the framework is free to include any additional parameters associated with the events used in a trigger statement.
Consequently, unlike the other operations, we associate the \emph{trigger} of a single event as the act of sending a message to any component in the system. Events are triggered in their specified order.

In terms of implementation, we can annotate our automaton with special \emph{active} states that represent \emph{trigger} operations.
These states do not transition on any events but simply instruct the framework to perform an action (send a message) on entry and immediately transition to the next state.
As an example, the statement \lstinline!expect e1 trigger e2 e3 expect e4! causes the test framework to send the messages associated with events \(e_2\) and \(e_3\) in between matching events \(e_1\) and \(e_4\).

\subsubsection{White-box Testing}

In addition to black-box testing by verifying external events using the constructs explained so far, Listing \ref{lst:inspect} shows the production for generating an \emph{inspect} statement.
The statement allows the tester to request an assertion over the internal state of the CUT at any point within the test execution.
This keyword prefixes a predicate function \(\alpha\) that takes as an argument the CUT and returns a boolean depending on whether the internal assertions run against the CUT succeeded or not.
On executing an \emph{inspect} statement, the framework pauses the execution of the test case and passes the instance of the CUT to the supplied predicate and resumes the test case when the predicate returns.
In the case of a unsuccessful inspection, where the returned value is false, the test case fails.

\begin{lstlisting}[label=lst:inspect, float=t!, caption={Inspecting Internal State}]
Body -> inspect cb
\end{lstlisting}

The idea of an inspection is that the tester puts the CUT into some known state by reproducing a set of conditions in the test environment and subsequently verifying the effects of internal events if any, performed by the CUT.
Producing and optionally verifying the required conditions to inspect the component internals is done using the same constructs for black-box testing such as \emph{trigger} and \emph{expect}.

A requirement is placed on the framework to offer some mechanism for performing inspections with the ability to pause the CUT's execution at runtime while its internal state is being inspected.
In view of components being scheduled nondeterministically, the framework must also enforce the happened-before relation \cite{lamport78} between all preceding actions, both internal and external, and the inspect operation.
As the components of message-passing systems are usually associated with a queue, messages to be handled by the CUT as a result of previous events may still be queued at the time of inspection.
Inspecting the CUT under such an incomplete environment results in an inaccurate verification of the CUT.
The framework must thus provide a transparent mechanism for ensuring that handlers for all events that happen-before the inspection at runtime have been executed and completed by the CUT before performing the inspection.

The \emph{inspect} statement can be incorporated into an automaton using the same technique as described in Section \ref{sec:trigger} where an active state is associated with each \emph{inspect} statement, instructing the framework to perform the inspection when it reaches that state.

\section{KompicsTesting - A Framework Implementation}
\label{sec:kompicstesting}
In this section we describe an implementation of the presented theoretical framework on Kompics. KompicsTesting is implemented on the reference implementation of Kompics \cite{kompicsImpl} which is written in Java. However, there are other implementations of Kompics in Python and Scala which are not considered in this paper. Thus we refer to the reference implementation simply as Kompics in this section.
We start with a description of the event symbols used to describe executions, followed by the architectural and design decisions that enable the implementation verify the described executions and enforce the operational semantics of the framework. Then we illustrate the DSL of the framework used for describing these executions according to the context-free grammar presented in the theoretical framework.

\subsection{The Alphabet of Executions}
\label{sec:alphabet}
As the main goal of KompicsTesting is to enable the tester describe a set of executions as a language, this section highlights the alphabet of the language from which executions are formed.
The theoretical framework described assumes the notion of an event and declares the set of all events to be the alphabet of the language.
Here, we make this idea concrete for our implementation.

Let \(E\) be the set containing all possible events observable in the environment of a test case in question, \(P\) the set of ports provided or required by all peer components in the environment including the CUT, and a set of literals \(D = \{in, out\}\) denoting the direction of events relative to the CUT.
As mentioned in Section \ref{sec:kompics}, Kompics provides a form of type system for events. We thus define our alphabet \(\Sigma\), the set of valid symbols for constructing executions, as the relation on the set \(E \times P \times D\) consisting of triples \((e,p,d)\) where \(e \in E, p \in P, d \in D\) and the port type of \(p\) allows events of type \(e\) in direction \(d\).
For example \((e, p, in) \in \Sigma\) if and only if the CUT provides port \(p\) and the port type of \(p\) allows \(e\) in the negative direction, or it requires \(p\) and the port type of \(p\) allows \(e\) in the positive direction.

The direction literals belonging to \(D\) are Java enums while event and port literals are provided as references to the test framework implementation.

\subsection{The Proxy Component}
The theoretical framework assumes that events incoming to and outoing from the CUT are not only observable by an implementation, but in operations like allow and drop, intercepted by the framework, since the CUT or peer components may be required not to receive such events.

KompicsTesting uses a special component as the \textit{Main} component in the test case called the \textit{proxy}, responsible for intercepting and observing events to and from the CUT.
This is enabled by Kompics' component heirarchy. As the first component to be created and bootstrapped by the Kompics runtime, the proxy acts as a composite component, in charge of creating the CUT and other peers as children components.

\begin{description}
\item[Internal Port Implementation in Kompics:]
As bidirectional interfaces, a port has both a negative and a positive side.
In the Kompics reference implementation, a port declaration (provided or required) \(\alpha\) by a component is internally represented by a pair of port references \((p, n)\) where \(p\) is the positive port and \(n\) is the negative port representing the two sides of \(\alpha\).
Both ports \(p\) and \(n\) are associated with an \emph{owner} which is either the component \(C\) that declares (provides or requires) the port or the parent component \(P\) of \(C\). On receiving an event on a port, the port's owner is scheduled for execution of its handlers.
For a port \(\alpha = (p,n)\) provided by a component \(C\), \(n\) is the inside port and owned by \(C\) while \(p\) is the outside port owned by parent \(P\). Conversely, for a port \(\alpha = (p,n)\) required by a \(C\), \(p\) is the inside port and owned by \(C\) while \(n\) is the outside port and owned by the parent \(P\).
\end{description}

Establishing a parent-child relationship between the proxy and all other components enables a sufficient degree of control over the events that go into and out of the child components.
Every peer component's port declaration assigns the outside port to the proxy.
Consequently, the proxy can subscribe handlers to these ports and listen for events on these ports. This, coupled with the fact that creating channels between any ports in the test environment must be done through the test framework, enables the proxy to completely isolate the CUT from other components, intercept events at its interfaces and only forward those which have not been explicitly disabled in the specification.

Consider a provided port \(\alpha = (p_\alpha,n_\alpha)\) by the CUT and a required port of the same type \(\beta = (p_\beta, n_\beta)\) by some peer component.
Say we connect both outside ports via a channel \(c\) from \(p_\alpha\) to \(n_\beta\) as would be done in a normal environment.
Handlers subscribed by the CUT for \(\alpha\) are done on \(n_\alpha\) and execute when an incoming event is received while outgoing events triggered by the CUT must go through \(p_\alpha\) and \(c\) before reaching the peer component.
If we subscribe intercepting handlers \(h_i\) to \(n_\alpha\) and \(h_o\) to \(p_\alpha\), we are able to observe both incoming and outgoing events respectively on port \(\alpha\).
However, since the handlers get executed at the same instance as those subscribed by the CUT, we are not able to optionally forward events as required by the framework, the CUT would have already received the event by the time our handlers execute.

The solution is to connect the proxy component directly to any peers that would normally be connected to the component while the CUT has no direct connections to any other components.
Due to Kompics' channel broadcast mechanism, peers are unaware of the source of received events and recipients of triggered events.
Thus by mirroring ports declared by the CUT on the proxy component, peers may be directly connected to the proxy while the CUT remains isolated.
In this case, a mirror port \(\delta = (p_\delta,n_\delta)\) of port \(\alpha\) would be created with \(c\) connected from \(p_\delta\) to \(n_\beta\).
Incoming events to are thus intercepted by subscribing \(h_i\) to \(n_\delta\) while \(h_o\) remains subscribed to \(p_\alpha\).
Handler \(h_o\) keeps track of the connected channels and components and manually forwards outgoing events when required.

\subsection{Scheduling}
Scheduling decisions must also be made in order to enfore the semantics of the framework.
So far we have described the process of intercepting and optionally forwarding events but not how to determine if an intercepted event should be forwarded.
KompicsTesting runs the automaton on a single control thread while other components except for the proxy are scheduled on a thread pool.
The proxy component uses a calling thread scheduler, executing its handlers immediately on the same thread of the component triggering the event on one of its ports.
This enables exactly those events that are going in and out of the CUT to be placed in an event queue as soon as they are triggered on any of the proxy's ports.
Since these components are scheduled independent from the automaton, the automaton can progress at its own pace and consume events from the queue when it needs to transition to some next state.
For example, a trigger state may add a new event to the queue while an expect state consumes an event from it.
Only when an event is consumed from a queue can it be \emph{handled}, that is forwarded to the recipient(s), as determined by the current state's transitions.

\subsection{A DSL for Building Specifications}
The simplicity of the presented context-free grammar allows us to implement our test specification scheme using a simple Java domain-specific modelling Language DSL, based on the builder pattern \cite{gamma93}.
It allows the tester to describe languages over our alphabet of events by calling API methods. 
The DSL uses a similar formation rule to the CFG, mapping keywords to API methods which can then be consumed by the tester as if they were generating a specification using the CFG.
An advantage of implementing a DSL directly in Java allows the tester to supply terminals of the CFG such as events, ports and predicate functions directly to the framework as references.

Generally, the API methods are named after an equivalent keyword in the CFG and where appropriate and take as arguments a representation of the literals, if any, as defined in the CFG.
For example, the \emph{inspect} statement requiring a predicate literal is implemented as a method \lstinline!inspect(Function<Component, Boolean>);! while the \emph{end} keyword is implemented as a method declared without any parameters.

Listing \ref{lst:apiSample} shows how the specification in Listing \ref{lst:cfgSample} would be written in KompicsTesting.
To avoid long-winding method calls, statements such as \emph{trigger} and \emph{expect} that prefix a sequence of literals are implemented as methods that take a single literal.
As a result, specifying multiple such literals in a sequence are done by calling the methods multiple times in succession, as shown using multiple calls to the \emph{expect} and \emph{allow} methods.
The trigger method only takes an event and a port reference but not direction. Since ports in Kompics are implemented as bidirectional port references, the direction is inferred from the supplied port reference.

\begin{lstlisting}[label=lst:cfgSample, float=t!, caption={Generating a specification from context-free grammar.}]
repeat 10
  allow $e_3$ $e_4$
body
  expect $e_1$ $e_2$
  trigger $e_5$
end
\end{lstlisting}

\begin{lstlisting}[basicstyle=\scriptsize\tt, float=t!, label=lst:apiSample, caption={Converting Listing \ref{lst:cfgSample} to a DSL.}]
testcontext.body()
  .repeat(10)
    .allow(e_3, p_3, in) 
    .allow(e_4, p_4, in)
  .body() 
    .expect(e_1, p_1, in)
    .expect(e_2, p_2, out)
    .trigger(e_5, p_5)
  .end();
\end{lstlisting}

\subsection{Creating and Executing a Test Case}
In KompicsTesting, the frst step of testing a component is by creating an instance of the \emph{TestContext} class by providing a component definition of the CUT to the test framework.
Listing \ref{lst:createTestCase} illustrates the process of creating and running a test case.
Line 1 creates a test context for the component under test having the component definition class \emph{CUT}. Line 2 retrieves a reference to the created CUT instance while Line 3 creates a peer component whose required port of type \emph{Message} is subsequently connected with the provided port of the CUT.
Once all such setup activities have been carried out, methods for for describing the execution of the CUT may be called (Line 6) and finally the \lstinline!boolean check();! method is called, running the testcase and returning true if the observed execution is a member of the language described by the test case (the created automaton ends up in an accepting state) and false otherwise.

The component definition is used by the framework to create an instance of the CUT for test execution, a reference to which is accessible to the tester throughout the specification.

\begin{lstlisting}[language=Java,basicstyle=\scriptsize\tt, label=lst:createTestCase, float=t!, caption={Creating and Executing a Test Case}]
TestContext<CUT> tc = TestContext.create(CUT.class);
Component cut = tc.getComponentUnderTest();
Component peer = tc.create(PeerComponent.class);
tc.connect(cut.getPositive(Message.class),
           peer.getNegative(Message.class));
//...
tc.check();
\end{lstlisting}

\subsection{Building an Automaton}
From the perspective of the tester, the DSL simply provides a means for describing the expected execution sequence. Once the \emph{check} method is called, the test framework takes control of the program.
It builds up an automaton using the techniques presented for each method called and concatenates them accordingly.
The final automaton is an \(\varepsilon\)-NFA, subsequently simulated by tracking a set of states that the automaton may possibly be in at anytime during test execution.
Simulation of the constructed automaton is done in parallel with the execution of the CUT. This enables the CUT to produce and handle events while the test framework matches these events and advances the automaton when required.

\subsection{Expecting Events}
As mentioned in section \ref{sec:alphabet}, event symbols are of the form \((e,p,d)\) where \(e\) is an event reference, \(p\) is a port reference and \(d\) is a direction enum.
Consequently, this triple must be specified to methods that add events to the described sequence.
Listing \ref{lst:expect} shows an example of expecting both ordered and unordered events.
Line 3 and 4 expects an incoming \(e_0\) and an outgoing \(e_1\) event in that specified order, creating two states with these events as transitions.
Line 5 through 8 expects both incoming \(e_2\) and outgoing \(e_3\) in any order, creating a single state, \ref{sec:orderingActions}, while the framework transitions to the final state only when both events have occurred.

\begin{lstlisting}[language=Java, float=t!, basicstyle=\scriptsize\tt, label=lst:expect, caption={Expecting Events}]
tc.
  body()
    .expect(e_0, p_0, in)
    .expect(e_1, p_1, out)
    .unordered()
      .expect(e_2, p_2, in)
      .expect(e_3, p_3, out)
    .end()
\end{lstlisting}

\lstset{
	language=Java,
	basicstyle=\scriptsize\tt
}

\subsubsection{Matching Events}
As KompicsTesting is written in Java, we would like to make sure that events are correctly matched according to the intentions of the tester, that is, we need some mechanism for defining the equality of events.
For example, in line 3 of listing \ref{lst:expect}, we say that we expect to match the event \(E_0 = (e_0, p_0, in)\) and having observed some event \(E_i = (e_i, p_i, d)\) at runtime, would like to verify that \(E_0\) is in fact \(E_i\).
Since directions and ports are internal to Kompics and the test framework, we can define our notion of equality transparently, albeit sensibly to the user. These are simply compared using the \lstinline!equals! method of \lstinline!Object!.
In other words, we need some user-defined notion of equality of the event references specified in executions.
Instead of forcing the tester to implement equals methods for each class with an instance that appears as an expected event, a comparator may be registered with the framework during test specification, for any class that may appear as an event, representing the users definition of equality for objects of that class.

Listing \ref{lst:comparatorpred} shows how this is accomplished via the setComparator method. The method can appear only within the initial header (the implicitly defined block header which will be explained in \ref{sec:blocksdsl}) of a test specification.
Say an event \((e_i, p_0, in)\) is received at the state with the transition labelled with the event specified in line 3, and that both \(e_0\) and the \(e_i\) belong to some class \(C\) which is assignable to the class \(E\) of the registered comparator in line 2, then this comparator is used to check for equality.
If no such comparator is registered for a class of events, then the framework defaults to using the equals method of the specified event.
Thus testers have the option of either providing a comparator for events or defining the equality of the event classes themselves.

Line 5 uses an explicit matching function for a class of events. The function provided is a predicate, in this case for class \(E\), that returns true if the event argument specified is matched and false otherwise.
This method provides allows the tester reuse the same predicate for describing a wider range of events instead of explicitly providing equivalent references of the expected events.
Consider the case of observing an event \((e_i, p_i, in)\) at some state \(q\) with the transition described in Line 5, and the class of \(e_i\) is assignable to \(E\), the specified class of the predicate.
Then the predicate is used to match \(e_i\).
Note that the behavior is not defined for a state with multiple transitions having predicates with the same signature as labels. Since such predicates could potentially match events of the same class so there is no way of correctly selecting an appropriate one. We thus declare such cases as a symptom of an ambiguous test specification and leave their behavior undefined.

\begin{lstlisting}[language=Java,basicstyle=\scriptsize\tt, label=lst:comparatorpred, float=t!, caption={Matching Events with Comparators and Predicates}]
tc
  .setComparator(E.class, comparatorE)
  .body()
    .expect(e_0, p_0, in)
    .expect(E.class, predicate, p_i, in)
\end{lstlisting}

\subsection{Request-Response}
KompicsTesting implements a variation of the more general request-response pattern described in section \ref{sec:reqres}.
Here the tester provides a function \(\alpha : E \to E\) mapping a given request to a response, which is subsequently triggered on some specified port.
Listing \ref{lst:requestResponse} illustrates this pattern.
Line 2 marks the starting point of the pattern while line 5 marks the end of the pattern.
The expect methods take in the class of the expected request, the outgoing port of the request (note that the direction is not specified since this is implicit as an outgoing request message), a second port where the provided response should be triggered and finally the mapping function that provides such a response.

The matching function is called with a received event and returns a response if the event is matched otherwise null.
Internally, the entire scheme is implemented as a single state that transitions to the next state only when all requests have been matched and the corresponding responses have been triggered.

\begin{lstlisting}[language=Java,basicstyle=\scriptsize\tt, label=lst:requestResponse, float=t!, caption={Request Response in KompicsTesting}]
tc.body()
  .expectWithMapper()
    .expect(R1.class, p_1, p_2, m1)
    .expect(R2.class, p_2, p_1, m2)
  .end()
\end{lstlisting}

\subsection{Creating Blocks}
\label{sec:blocksdsl}
Blocks declared within the \emph{repeat} and \emph{end} keywords are represented by their method variants.
Listing \ref{lst:repeat} shows the pattern for creating blocks within KompicsTesting.
As every event must belong to a block, an entire test case is always run in a block that is repeated exactly once.
This block is not explicitly created by the tester, in the sense that the corresponding repeat and end methods are not called. The \lstinline!body! method however, must still be called before using non-header methods as shown in line 1.
Two blocks are explicitly created by the tester; the first spans Line 2 through 5 and triggers some \(e_0\) on the specified port 3 times, while the second block creates a Kleene closure block that matches as many incoming \(e_1\) events as possible on port \(p_1\).
Blocks with a fixed number of iterations are implemented using a loop counter that is decremented for each traversal of the state graph representing the block. The graph is exited when the counter reaches zero. This avoids the need to create a large number of states for large sizes of \(n\).

\begin{lstlisting}[language=Java,basicstyle=\scriptsize\tt, label=lst:repeat, float=t!, caption={Creating Blocks in KompicsTesting}]
tc.body()
  .repeat(3)
  .body()
      .trigger(e_0, p_0)
  .end()
  .repeat()
  .body()
      .expect(e_1, p_1, d_1)
  .end()
\end{lstlisting}

\subsection{Conditional Statements}
Listing \ref{lst:conditional} shows the pattern for creating a conditional statement in KompicsTesting. In this example, lines 3 and 4 comprise the \emph{either} branch and expects two events while lines 6 through 10 comprise the \emph{or} branch of the statement, expecting any number of the event sequence described in the body of the repeat block.
Similar to the Kleene closure block equivalent, this statement creates a NFA.
In this example, the start state contains a transitions on \(E = (e_1, p_1, d_1)\) to both the states representing statements in lines 4 and 9, from where on the automaton diverges.

\begin{lstlisting}[language=Java,basicstyle=\scriptsize\tt, label=lst:conditional, float=t!, caption={Conditional Statements}]
tc.body()
  .either()
    .expect(e_1, p_1, d_1)
    .expect(e_2, p_2, d_2)
  .or()
    .repeat()
    .body()
        .expect(e_3, p_3, d_3)
        .expect(e_4, p_4, d_4)
    .end()    
  .end()
\end{lstlisting}

\subsection{Interactive Testing}
Listing \ref{lst:interactive} shows a demonstration of the \emph{trigger} and \emph{inspect} methods for a specification.
This example interleaves the trigger with the inspect methods that assert any changes to the component state made due to the previous trigger. Here the reference supplied as an argument to the inspect method is an instance of a Function$<$C, Boolean$>$ where $<$C extends ComponentDefinition$>$ represents the type of the component definition of the CUT.

\begin{lstlisting}[language=Java,basicstyle=\scriptsize\tt, label=lst:interactive, float=t!, caption={Interactive Tests}]
tc.body()
  .trigger(e_0, p_0)
  .inspect(predicate)
  .trigger(e_1, p_1)
  .inspect(predicate)
\end{lstlisting}

As the \emph{trigger} operation is used to simulate components sending events across channels, a triggered event in KompicsTesting is simply published on any connected channels by triggering on the given port.

The \emph{inspect} operation requires that all previous events that happened-before the operation must have been handled by the CUT before the inspection can commence.
To accomplish this, on entry to the inspect state, KompicsTesting verifies that the work count, the amount of pending events and events currently being handled, is zero otherwise it waits until this becomes true, signalling that all events have been handled by the CUT.

\subsection{Ambiguous Test Specifications}
The introduction of interactive operations like trigger and inspect, while useful tools for testing, come with caveats when writing test cases with inherent nondeterminism.
Consider a Kleene closure that does not describe any sequence as shown in \ref{lst:ambiguousKleene} and a conditional statement with both branches beginning with interactive operations as shown in \ref{lst:ambiguousCondition}.
In the first case there is no way of knowing how many times the trigger operation should be performed since the block does not define any clear entry or exit conditions.
The second case does not provide a way to accurately choose between the \emph{either} and \emph{or} branches since they both do not have entry conditions.
Now consider listing \ref{lst:unambiguousCondition} without a clear entry condition for the \emph{either} branch but an entry condition for the \emph{or} branch.
From the start state, the automaton may consult the event queue and only if it does not consume the event \((e_1, p_1, d_1)\) successfully does it traverse the either branch.

The testers intent can not be inferred by the framework from specifications such as those in listings \ref{lst:ambiguousKleene} and \ref{lst:ambiguousCondition} and in fact, the intent of the first case is questionable to begin with.
Such specifications are deemed ambiguous and the behavior of the framework is left undefined.
Listing \ref{lst:unambiguousCondition} on the other hand is not ambiguous.

\begin{lstlisting}[language=Java,basicstyle=\scriptsize\tt, label=lst:ambiguousKleene, float=t!, caption={A Kleene block that does not describe any execution}]
tc.body()
  .repeat()
      .trigger(e_0, p_0)
  .body()
  .end()
\end{lstlisting}

\begin{lstlisting}[language=Java,basicstyle=\scriptsize\tt, label=lst:ambiguousCondition, float=t!, caption={A Conditional statement with no clear entry conditions}]
tc.body()
  .either()
    .trigger(e_0, p_0)
  .or()
    .trigger(e_1, p_1)
  .end()
\end{lstlisting}

\begin{lstlisting}[language=Java,basicstyle=\scriptsize\tt, label=lst:unambiguousCondition, float=t!, caption={A Conditional statement with an entry condition for the \emph{or} branch}]
tc.body()
  .either()
    .trigger(e_0, p_0)
  .or()
    .expect(e_1, p_1, d_1)
  .end()
\end{lstlisting}

\subsection{Accepting Sequences and Termination}
A test case is successful when the current set of states contains a final state and no other events are received as that would take the automaton a next state.
Since the processing of events and simulation of the automaton is done at real-time with the execution of components, it may sometimes become problematic to predict when no new events are expected or when the test case should terminate. The unpredictable scheduling and asynchronous behavior of components are the reasons for this.

Listing \ref{lst:timeout} illustrates both cases. The tester initially expects to receive any number of the event specified at line 4. However the automaton may prematurely consult an empty event queue and either immediately fail with an end-of-stream error, or try to wait, potentially indefinitely, for the event.
Lines 3 through 6 on the other hand expect any number of the event specified on line 5 for successful termination.
Here, the automaton may choose to transition to an accepting state if the event queue is empty but this behavior would not be correct if there are some events still in transit.
To resolve this, KompicsTesting includes a default timeout, configurable by the tester that specifies how long the framework should wait for an event to occur.
The responsiblity of enforcing a correct verification of the component in such cases is thus placed on the tester.

\begin{lstlisting}[language=Java,basicstyle=\scriptsize\tt, label=lst:timeout, float=t!, caption={A test without timeout for events}]
tc.body()
  .expect(e_0, p_0, d_0)
  .repeat()
  .body()
    .expect(e_1, p_1, d_1)
  .end()
\end{lstlisting}

\section{Related Work}
\label{sec:relatedwork}

This paper presents an approach for verifying the correctness of message passing systems.
Another strategy for accomplishing the same goal is through the use of model checking, a formal method used to verify the conformance of a given model of a system to its intended specification, containing properties which the system must satisfy.
Model checking methods use state-space exploration \cite{Clarke:2001:MC:778522.778533}, and can instill more confidence into a system implementation by automatically enumerating and exhaustively exploring the state-space of the system in order to find errors.
If all paths in the state-space have successfully been executed, then the system is said to be correct.
However, generating the state-space of a system can be an expensive process and in the presence of concurrency as is typical of message passing systems, it is not uncommon for a exponentially larger or infinite number of states to be required in order to perform properly verify the system due to the number of possible interleavings in the initial state space, a problem known as state-space explosion.
Techniques such as partial-order reduction \cite{Maiya:2016:POR:2945480.2945490}, \cite{Bokor:2011:SDS:2190078.2190164} and dynamic partial-order reduction exist to \cite{Lauterburg:2010:EOH:2128562.2128592} try to mitigate this problem by analysing opportunities for reducing the number of explored states but do not always succeed managing the problem.

Previously, model checking methods have been applied to testing of message passing systems. Systems like Basset \cite{lauterburg2009framework}, Jacco \cite{Zakeriyan:2015:JME:2824815.2824819} and McErlang \cite{fredlund2007mcerlang} have been developed for actor systems.
Basset provides a model checker for actor systems that compile to Java bytecode, using Java PathFinder JPF \cite{visser2003model}, an environment for preforming model checking, verification and analysis of Java programs.
In order to efficiently explore a Java actor system, it avoids exploring low-level and library code by concentrating on message scheduling related errors by providing a transparent interface to the actor application being tested while replacing the underlying library code with equivalent code that is easier to explore.
It uses dynamic partial-order reduction as an optimization technique to prune the number of explored states. However, its message scheduling assumptions leads to false negatives in generating explorable states leading to large state spaces of programs.
Its reliance on the JPF platform also leads to inefficient test execution time.
Jacco, improves on the Basset system by implementing a new scheduling approach and technique for reducing the state space of the actor system in order to reduce the time taken for model checking. Coupled with an architecture that abandons the reliance on the JPF platform, it shows better performance in both execution time and generated state space of programs compared to Basset.
McErlang is a model checker for Erlang programs. Being developed in Erlang itself, it facilitates rapid prototyping and testing of Erlang programs and supports a substantial part of the programming language including the distributed parts.
It allows the user to encode correctness properties of the system as automata specified in Erlang while the model checker checks the system against the specified properties.
However, it uses state-space exploration for systematic checking of the system against the specification and suffers from the state-explosion problem especially in the face of nondeterminism.

Our approach presents a platform independent framework and foregoes the use state-space generation and exploratory techniques, depending solely on the tester to specify the particular state-space to be exercised.
Thus it does not run into problems like state-explosion that can make verification infeasible. However, the possibility of offering full test coverage of the component depends solely on the tester.
Our approach also focuses solely on verifying a single component in the system as opposed to system-wide verification as in the mentioned tools. As a consequence, a tester is unable to verify global properties of the system using a tool based on our approach.

The Akka TestKit tool \cite{akkatestkit} on the other hand provides a platform for performing unit and integration testing on actor systems based on the Akka framework at various level of granularity.
In terms of unit testing, it allows the tester perform synchronous tests on specific units of code without involving the actor model, in addition to performing unit tests on a higher level of abstraction using single actors as units.
In the latter case, a tester is able to send messages to an actor under test and subsequently verify responses in a synchronous manner. In between message exchanges, the tester is able to access the actors internal state in order to perform white-box testing similarly to our approach.
It also allows the user test that an incoming sequences of events is processed correctly, even in the face of nondeterminism, leading to reordering of events. However, such tests can only be performed using multiple actors to generate the stream of events.
Unlike our approach that allows an interactive mechanism for generating streams of events.
Since it does not use an approach of describing the entire sequence, verifying outgoing events is done manually by the tester, for example setting up a recipient component that asserts the outgoing events.

\section{Conclusions}
\label{sec:conclusion}

In this paper we presented a language based approach to unit testing message passing systems as well as a prototype implementation of this approach for the reference implementation of the Kompics component framework.
In the first part of the paper, we presented the rationale for treating sequences of events incoming to, and outgoing from a messaging component as a language and how such a language could be used to efficiently verify the behavior of the component using blackbox and white box based testing techniques.
This was done by presenting a theoretical framework for testing components by describing its expected sequence of events and interactively introducing events into this sequence in order to produce stimuli during test execution.
We showed how a tester may describe deterministic and nondeterministic sequences as well as how an implementing framework may cope with the feasibility of verifying nondeterministic sequences.

The second part of the paper presented KompicsTesting as a prototype implementation of the theoretical framework. We discussed the design choices and internal implementation of the prototype and illustrated the provided DSL for describing sequences. We showed how the DSL may be used to specify a wide variety of sequences modelling the behavior of components in Kompics.

\bibliographystyle{jfp}
\bibliography{Kompics}  

\begin{thebibliography}{}

\bibitem[\protect\citename{Akka, }n.d.]{akkatestkit}
Akka.
\newblock {\em Testing actor systems}.

\bibitem[\protect\citename{Alagar \& Periyasamy, }2011]{alagar11}
Alagar, V.~S., \& Periyasamy, K. (2011).
\newblock {\em Extended finite state machine}.
\newblock London: Springer London.
\newblock Pages  105--128.

\bibitem[\protect\citename{Arad, }2013]{Arad2013a}
Arad, Ci Cosmin~Ionel. (2013).
\newblock {\em {Programming Model and Protocols for Reconfigurable Distributed
  Systems}}.
\newblock Ph.D. thesis, KTH - Royal Institute of Technology, Stockholm.

\bibitem[\protect\citename{Arad, }2008]{kompicsImpl}
Arad, Cosmin. (2008).
\newblock {\em Kompics project}.

\bibitem[\protect\citename{Armstrong, }2003]{Armstrong2003}
Armstrong, Joe. (2003).
\newblock {Making reliable distributed systems in the presence of software
  errors}.
\newblock  295.

\bibitem[\protect\citename{Autebert {\em et~al.}\relax,
  }1997]{Autebert97context-freelanguages}
Autebert, Jean-Michel, Berstel, Jean, \& Boasson, Luc. (1997).
\newblock Context-free languages and push-down automata.
\newblock {\em Pages  111--174 of:} {\em Handbook of formal languages}.
\newblock Springer.

\bibitem[\protect\citename{Bokor {\em et~al.}\relax,
  }2011]{Bokor:2011:SDS:2190078.2190164}
Bokor, Peter, Kinder, Johannes, Serafini, Marco, \& Suri, Neeraj. (2011).
\newblock Supporting domain-specific state space reductions through local
  partial-order reduction.
\newblock {\em Pages  113--122 of:} {\em Proceedings of the 2011 26th ieee/acm
  international conference on automated software engineering}.
\newblock ASE '11.
\newblock Washington, DC, USA: IEEE Computer Society.

\bibitem[\protect\citename{Chomsky, }1956]{chomsky1956three}
Chomsky, Noam. (1956).
\newblock Three models for the description of language.
\newblock {\em Ire transactions on information theory}, {\bf 2}(3), 113--124.

\bibitem[\protect\citename{Clarke \& Schlingloff,
  }2001]{Clarke:2001:MC:778522.778533}
Clarke, Edmund~M., \& Schlingloff, Bernd-Holger. (2001).
\newblock Handbook of automated reasoning.
\newblock Amsterdam, The Netherlands, The Netherlands: Elsevier Science
  Publishers B. V.

\bibitem[\protect\citename{Fredlund \& Svensson, }2007]{fredlund2007mcerlang}
Fredlund, Lars-{\AA}ke, \& Svensson, Hans. (2007).
\newblock Mcerlang: a model checker for a distributed functional programming
  language.
\newblock {\em Pages  125--136 of:} {\em Acm sigplan notices},  vol. 42.
\newblock ACM.

\bibitem[\protect\citename{Gamma {\em et~al.}\relax, }1993]{gamma93}
Gamma, Erich, Helm, Richard, Johnson, Ralph, \& Vlissides, John. (1993).
\newblock {\em Design patterns: Abstraction and reuse of object-oriented
  design}.
\newblock Berlin, Heidelberg: Springer Berlin Heidelberg.
\newblock Pages  406--431.

\bibitem[\protect\citename{Hopcroft \& Ullman, }1979]{hopcroftullman79}
Hopcroft, John~E., \& Ullman, Jeffrey~D. (1979).
\newblock {\em Introduction to automata theory, languages, and computation}.
\newblock Addison-Wesley Publishing Company, Inc.

\bibitem[\protect\citename{{IEEE}, }2010]{IEEESEVocab2010}
{IEEE}. (2010).
\newblock {\em {ISO/IEC/IEEE 24765: Systems and Software Engineering -
  Vocabulary}}.
\newblock Tech. rept.

\bibitem[\protect\citename{Kleene, }1951]{kleene1951representation}
Kleene, Stephen~Cole. (1951).
\newblock {\em Representation of events in nerve nets and finite automata}.
\newblock Tech. rept. DTIC Document.

\bibitem[\protect\citename{Lamport, }1978]{lamport78}
Lamport, Leslie. (1978).
\newblock Time, clocks, and the ordering of events in a distributed system.
\newblock {\em Commun. acm}, {\bf 21}(7), 558--565.

\bibitem[\protect\citename{Lauterburg {\em et~al.}\relax,
  }2009]{lauterburg2009framework}
Lauterburg, Steven, Dotta, Mirco, Marinov, Darko, \& Agha, Gul. (2009).
\newblock A framework for state-space exploration of java-based actor programs.
\newblock {\em Pages  468--479 of:} {\em Proceedings of the 2009 ieee/acm
  international conference on automated software engineering}.
\newblock IEEE Computer Society.

\bibitem[\protect\citename{Lauterburg {\em et~al.}\relax,
  }2010]{Lauterburg:2010:EOH:2128562.2128592}
Lauterburg, Steven, Karmani, Rajesh~K., Marinov, Darko, \& Agha, Gul. (2010).
\newblock Evaluating ordering heuristics for dynamic partial-order reduction
  techniques.
\newblock {\em Pages  308--322 of:} {\em Proceedings of the 13th international
  conference on fundamental approaches to software engineering}.
\newblock FASE'10.
\newblock Berlin, Heidelberg: Springer-Verlag.

\bibitem[\protect\citename{Lynch \& Tuttle, }1989]{lynch89}
Lynch, Nancy~A., \& Tuttle, Mark~R. (1989).
\newblock An introduction to input/output automata.
\newblock {\em Cwi quarterly}, {\bf 2}, 219--246.

\bibitem[\protect\citename{Mackinnon {\em et~al.}\relax, }2000]{mackinnon00}
Mackinnon, Tim, Freeman, Steve, \& Craig, Philip. (2000).
\newblock Endo-testing: Unit testing with mock objects.
\newblock {\em extreme programming and flexible processes insoftware
  engineering - xp2000}.

\bibitem[\protect\citename{Maiya {\em et~al.}\relax,
  }2016]{Maiya:2016:POR:2945480.2945490}
Maiya, Pallavi, Gupta, Rahul, Kanade, Aditya, \& Majumdar, Rupak. (2016).
\newblock Partial order reduction for event-driven multi-threaded programs.
\newblock {\em Pages  680--697 of:} {\em Proceedings of the 22nd international
  conference on tools and algorithms for the construction and analysis of
  systems - volume 9636}.
\newblock New York, NY, USA: Springer-Verlag New York, Inc.

\bibitem[\protect\citename{Mateescu \& Salomaa,
  }1997]{Mateescu:1997:FLI:267846.267847}
Mateescu, Alexandru, \& Salomaa, Arto. (1997).
\newblock Handbook of formal languages, vol. 1.
\newblock New York, NY, USA: Springer-Verlag New York, Inc.

\bibitem[\protect\citename{Rabin \& Scott,
  }1959]{Rabin:1959:FAD:1661907.1661909}
Rabin, M.~O., \& Scott, D. (1959).
\newblock Finite automata and their decision problems.
\newblock {\em Ibm j. res. dev.}, {\bf 3}(2), 114--125.

\bibitem[\protect\citename{Raynal \& Singhal, }1996]{raynal96}
Raynal, Michel, \& Singhal, Mukesh. (1996).
\newblock Logical time: Capturing causality in distributed systems.
\newblock {\em Computer}, {\bf 29}(2), 49--56.

\bibitem[\protect\citename{Visser {\em et~al.}\relax, }2003]{visser2003model}
Visser, Willem, Havelund, Klaus, Brat, Guillaume, Park, SeungJoon, \& Lerda,
  Flavio. (2003).
\newblock Model checking programs.
\newblock {\em Automated software engineering}, {\bf 10}(2), 203--232.

\bibitem[\protect\citename{Wyatt, }2013]{Wyatt:2013:AC:2663429}
Wyatt, Derek. (2013).
\newblock {\em {Akka Concurrency}}.
\newblock USA: Artima Incorporation.

\bibitem[\protect\citename{Zakeriyan {\em et~al.}\relax,
  }2015]{Zakeriyan:2015:JME:2824815.2824819}
Zakeriyan, Arvin, Khamespanah, Ehsan, Sirjani, Marjan, \& Khosravi, Ramtin.
  (2015).
\newblock Jacco: More efficient model checking toolset for java actor programs.
\newblock {\em Pages  37--44 of:} {\em Proceedings of the 5th international
  workshop on programming based on actors, agents, and decentralized control}.
\newblock AGERE! 2015.
\newblock New York, NY, USA: ACM.

\end{thebibliography}

\label{lastpage}

\end{document}